\documentclass[twocolumn,showpacs,preprintnumbers,amsmath,amssymb]{revtex4}
\usepackage{graphicx}
\usepackage{dcolumn}
\usepackage{bm}% bold math

\def \beqn {\begin{equation}}
\def \bfig {\begin{figure}}
\def \btab {\begin{table}}
\def \eeqn {\end{equation}}
\def \efig {\end{figure}}
\def \etab {\end{table}}

\def \cuscn{$\kappa$-(ET)$_2$Cu(NCS)$_2$}

\def \fs{Fermi surface}

\def \Bc2{$B_{\rm c2}$}
\def \deg{$^\circ$}

\def \pf6{(TMTSF)$_2$PF$_6$}
\def \Ef{$E_{\rm F}$}

\def \bz{Brillouin zone}
\def \rhozz{$\rho_{zz}$}

\newcommand{\half}{{\frac{1}{2}}}

\newcommand{\chapstyle}{\sffamily}
\newcommand{\Fref}[1]{Figure~{\ref{#1}}}
\newcommand{\Eref}[1]{Equation~{\ref{#1}}}
\newcommand{\Tref}[1]{Table~{\ref{#1}}}
\newcommand{\Cref}[1]{Chapter~{\ref{#1}}}
\newcommand{\Sref}[1]{Section~{\ref{#1}}}

\def \cuscnfs{
\begin{figure}[tbp]
\centering
\includegraphics[height=7cm]{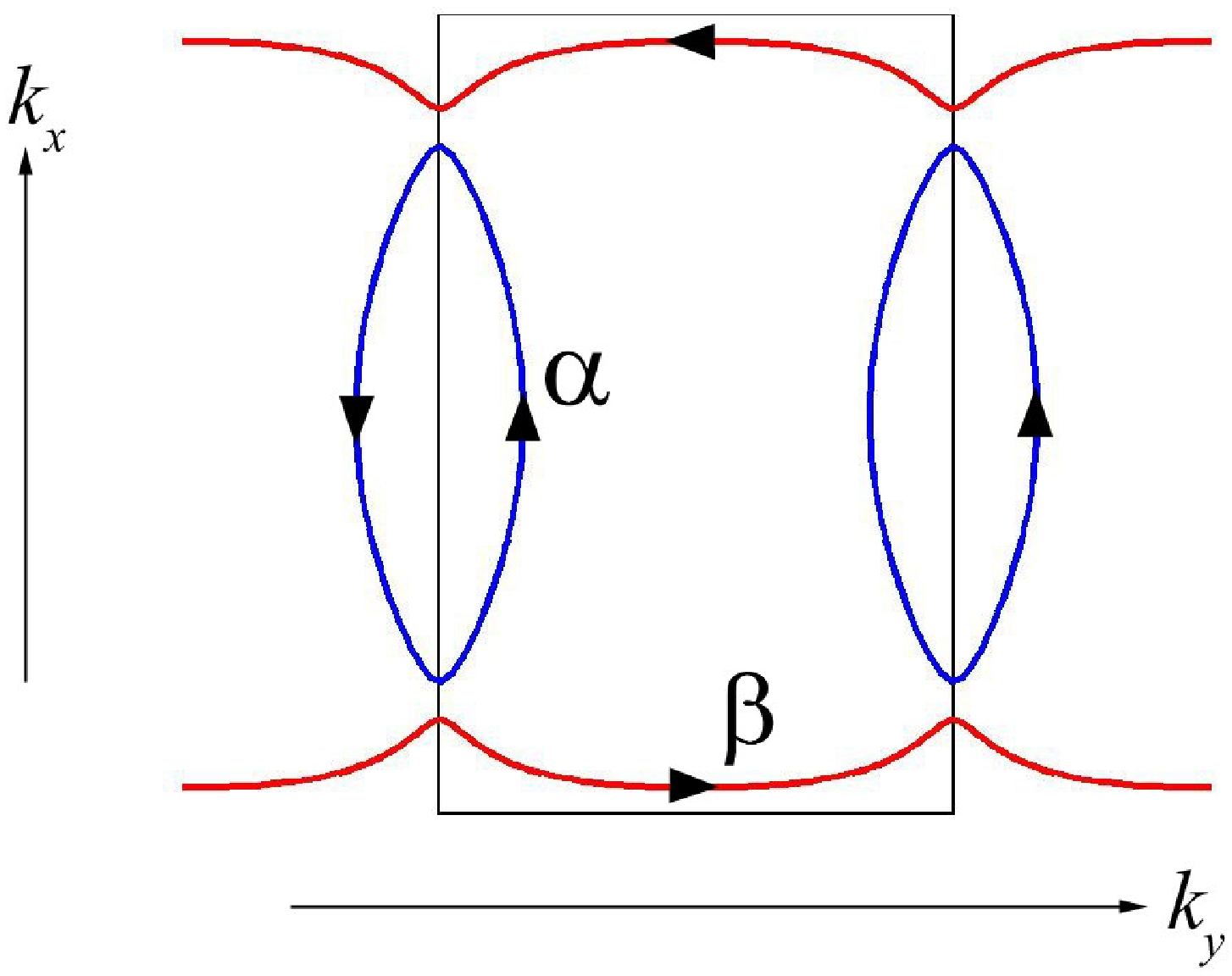}
\includegraphics[height=7cm]{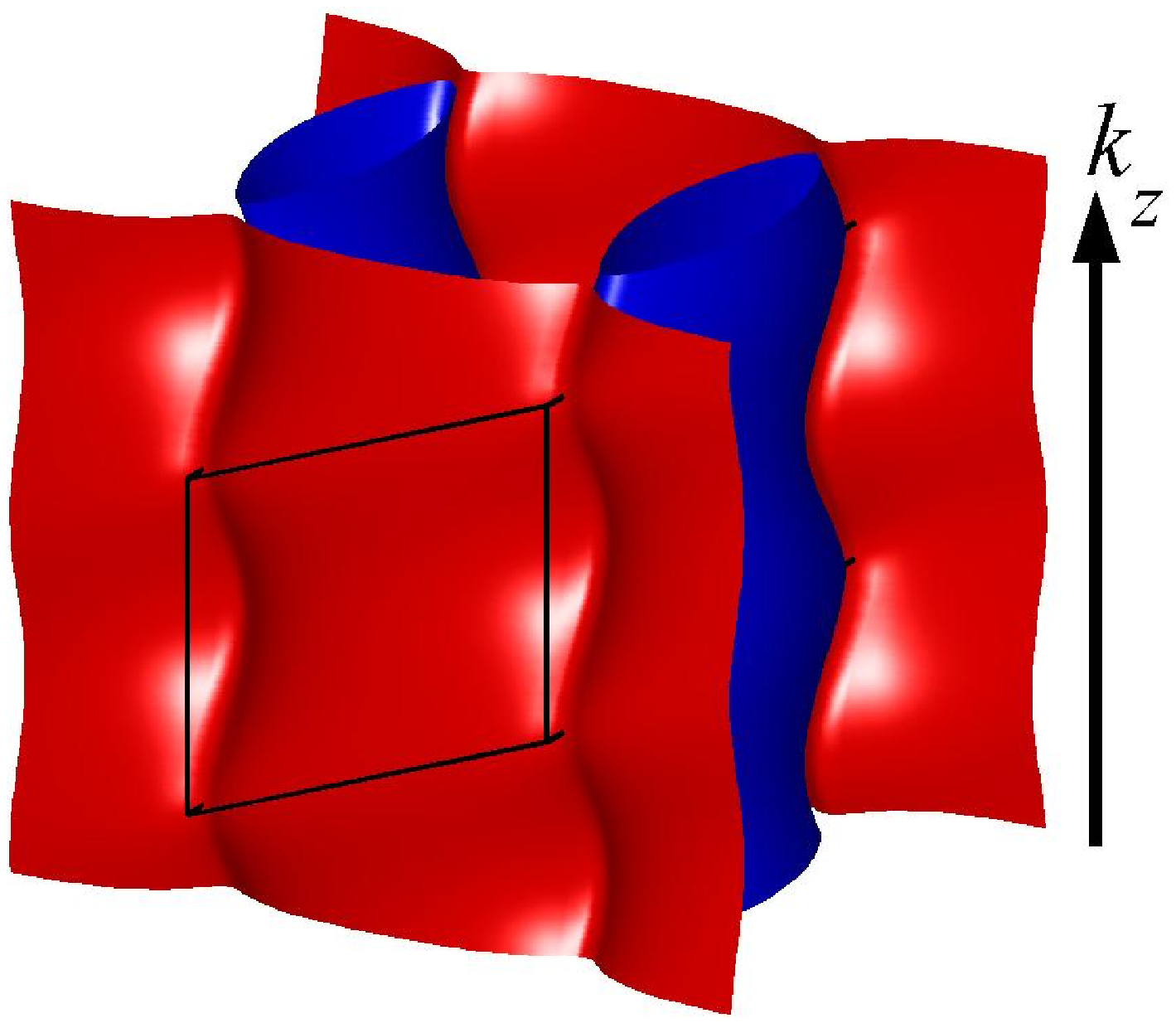}
\caption{Top: Cross-section of the \fs~of \cuscn~in the $k_xk_y$-plane.
The shape is defined by three transfer integrals in the highly conducting
layers (see \Eref{effectivedimer}). Bottom: The same \fs~in three dimensions. The interlayer
warping is defined by the transfer integral $t_a$ (see \Eref{cuscndisp}), and is
exaggerated for clarity.} \label{cuscnfs}
\end{figure}
}

\def \typicalsweep{
\begin{figure*}[tbp]
\centering
\includegraphics[height=12cm]{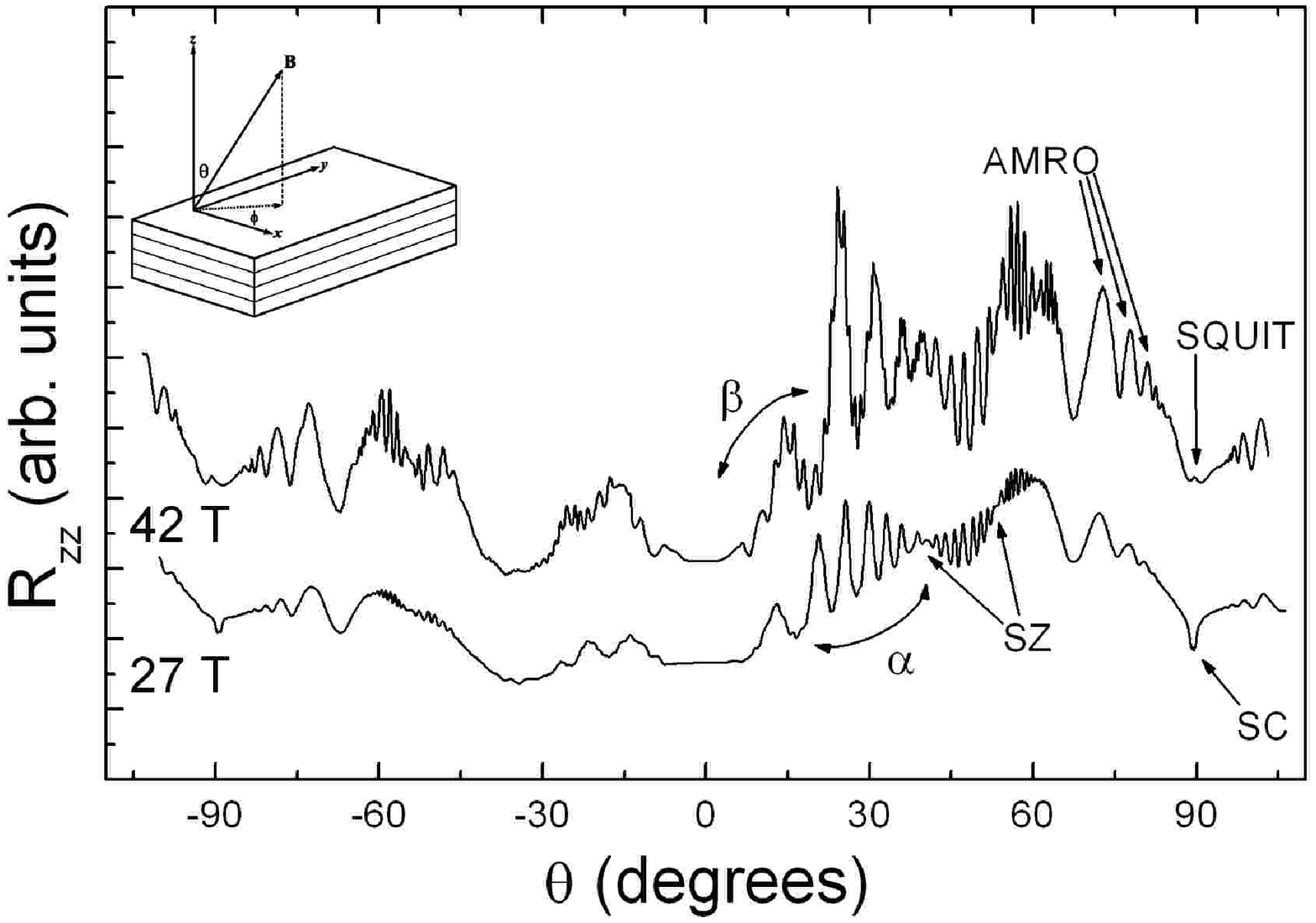}
\caption{Typical $\theta$-dependence of the magnetoresistance of
\cuscn. The data shown is for a hydrogenated sample at 490~mK,
$\phi=149$\deg (where $\phi$ is the azimuthal angle), 27~T (lower) 
and 42~T (upper). The data have been offset for clarity. Some representative features are indicated; SdH oscillations due to the Q2D pockets ($\alpha$) and the breakdown orbit ($\beta$); spin-zeroes in the SdH amplitudes (SZ);
the onset of the superconducting transition (SC); angle-dependent
magnetoresistance oscillations (AMRO), whose positions are field
independent; and the resistive peak in the presence of an exactly
in-plane magnetic field (In-plane Peak). The inset diagram is included to illustrate the measurement geometry.} \label{typicalsweep}
\end{figure*}
}

\def \lkfits{
\begin{figure}[tbp]
\centering
\includegraphics[height=6cm]{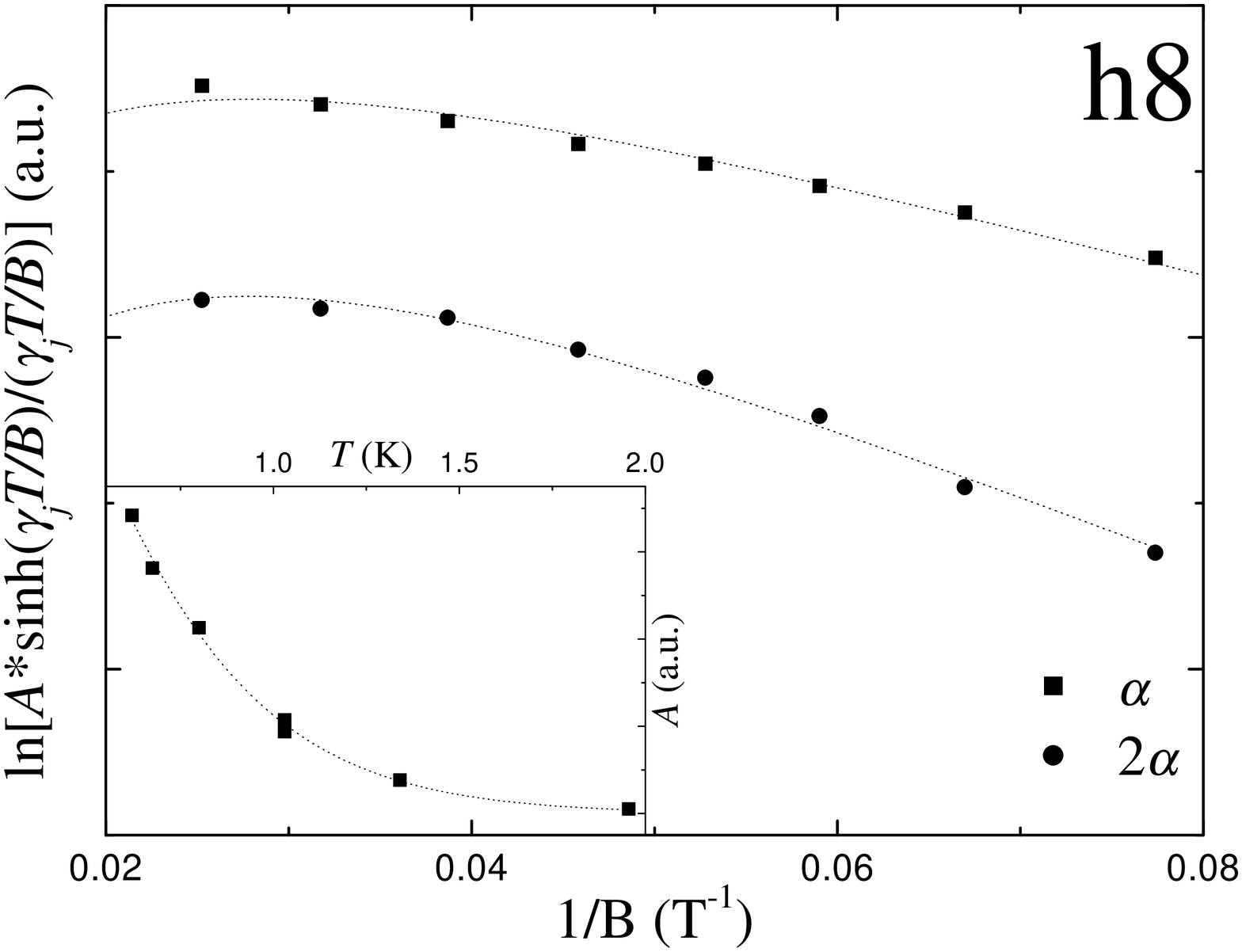}
\includegraphics[height=6cm]{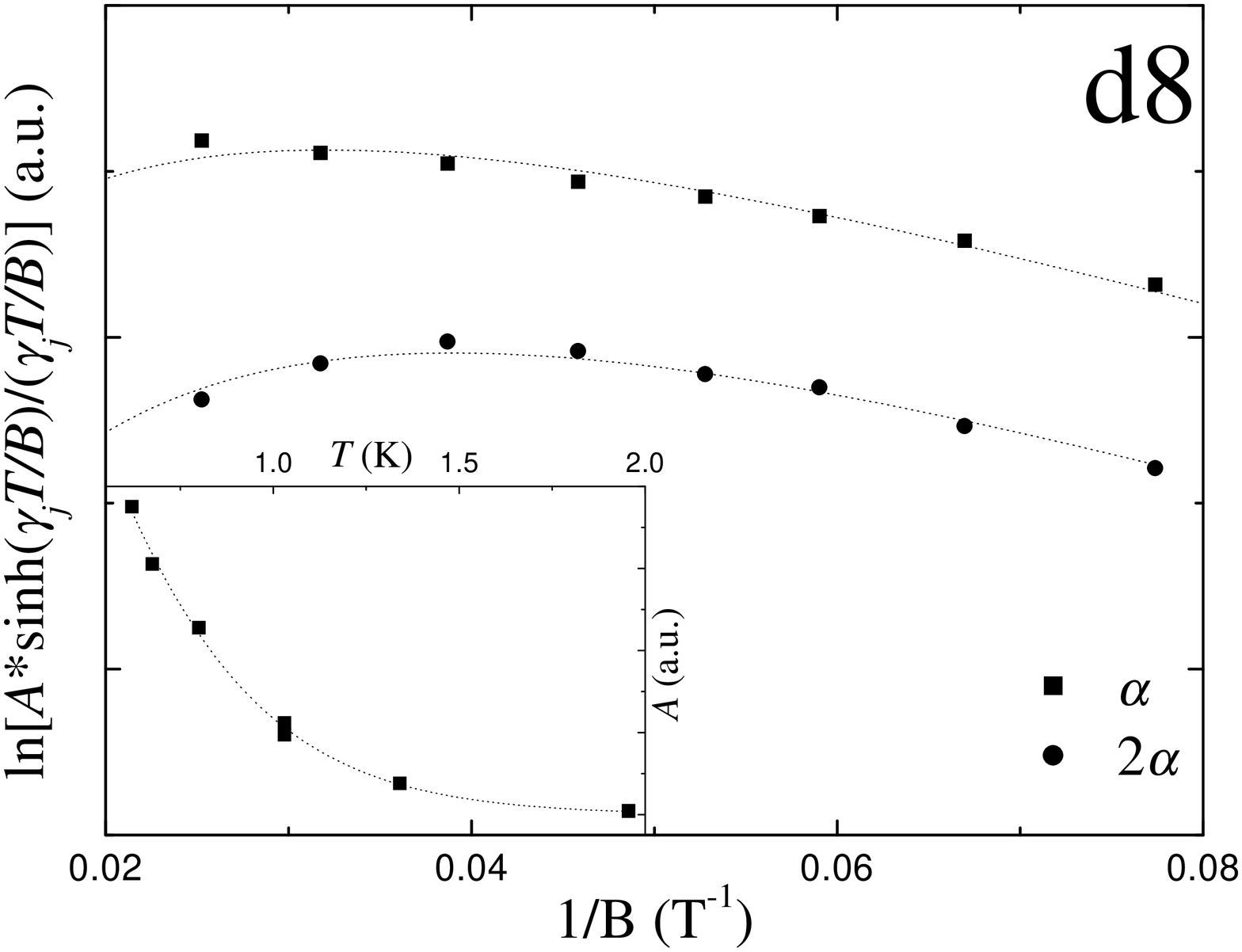}
\caption{The result of fitting the Fourier amplitude, $A$, of the
$h8$ (top) and $d8$ (bottom) $\alpha$-frequency (squares) and its
second harmonic (circles) to the two-dimensional Lifshitz-Kosevich
formula at constant temperature, over a field range of 12 -- 44~T,
using the technique outlined in the text and
Reference~\cite{harrison96mb}. This kind of analysis yields values for the scattering time and the magnetic breakdown field. The insets show the result of
fitting the temperature dependence of the $h8$ (top) and $d8$
(bottom) $\alpha$-frequency amplitude over a constant (low-)field
interval. The $h8$ and $d8$ data at each temperature were taken simultaneously and the fact that the insets are similar is an indication that the effective masses of the two isotopic substitutions are also similar.}
\label{lkfits}
\end{figure}
}

\def \cuscnsz{
\begin{figure}[tbp]
\centering \includegraphics[height=5.8cm]{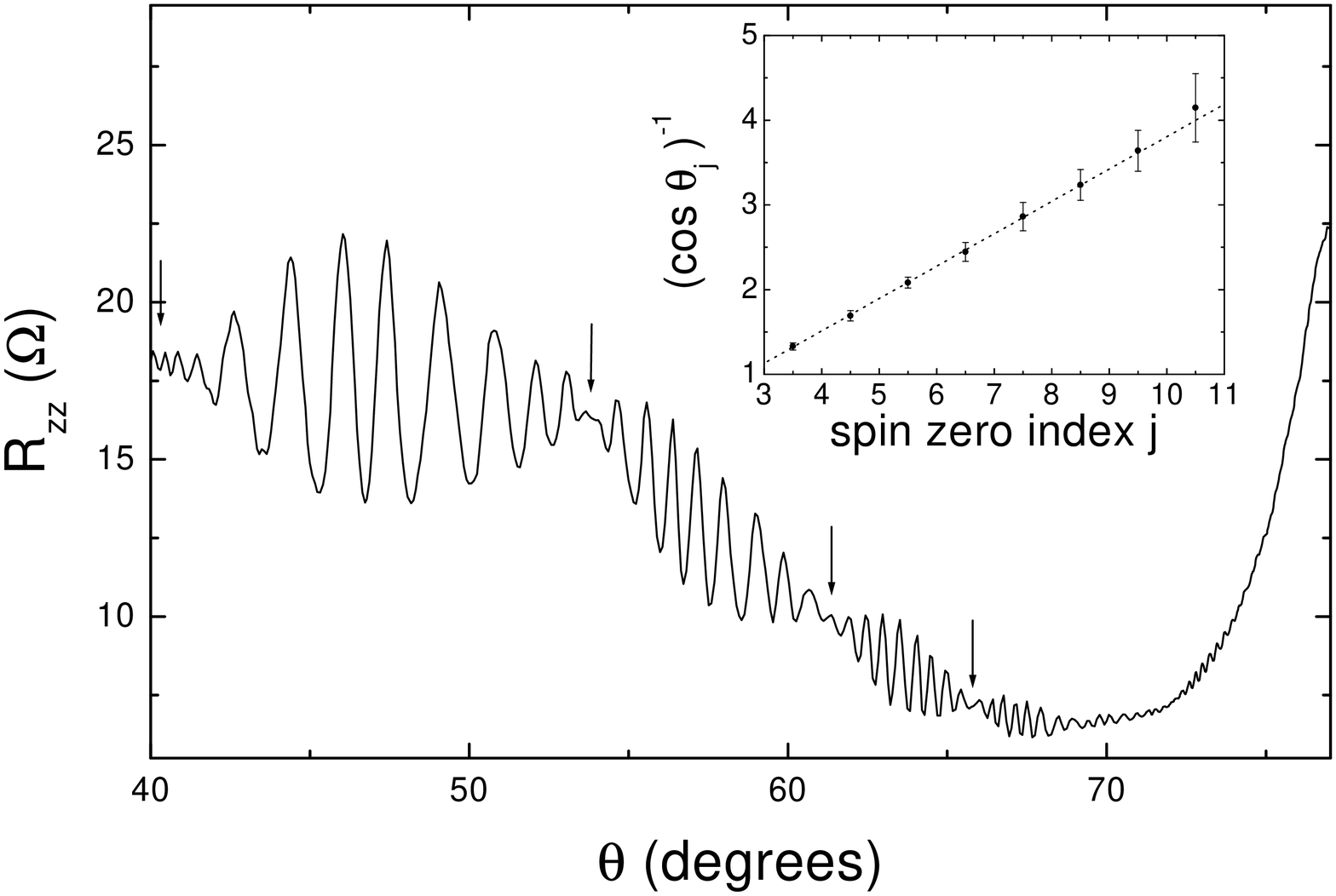}
\caption{The spin-zero effect as seen in a typical
$\theta$-dependence of the magnetoresistance of \cuscn. The data
shown is for an $h8$ sample at 590~mK, $\phi=90.8$\deg, and 27~T.
The spin-zero angles, $\theta_j$ are indicated by arrows. The data
points in the inset are a summary of the $\theta_j$ observed at a
number of different azimuthal angles, magnetic fields, and in two
different single crystal samples. The broken line is a straight
line fit to these data.} \label{cuscnsz}
\end{figure}
}

\def \cuscnsplit{
\begin{figure}[tbp]
\centering \includegraphics[height=9cm]{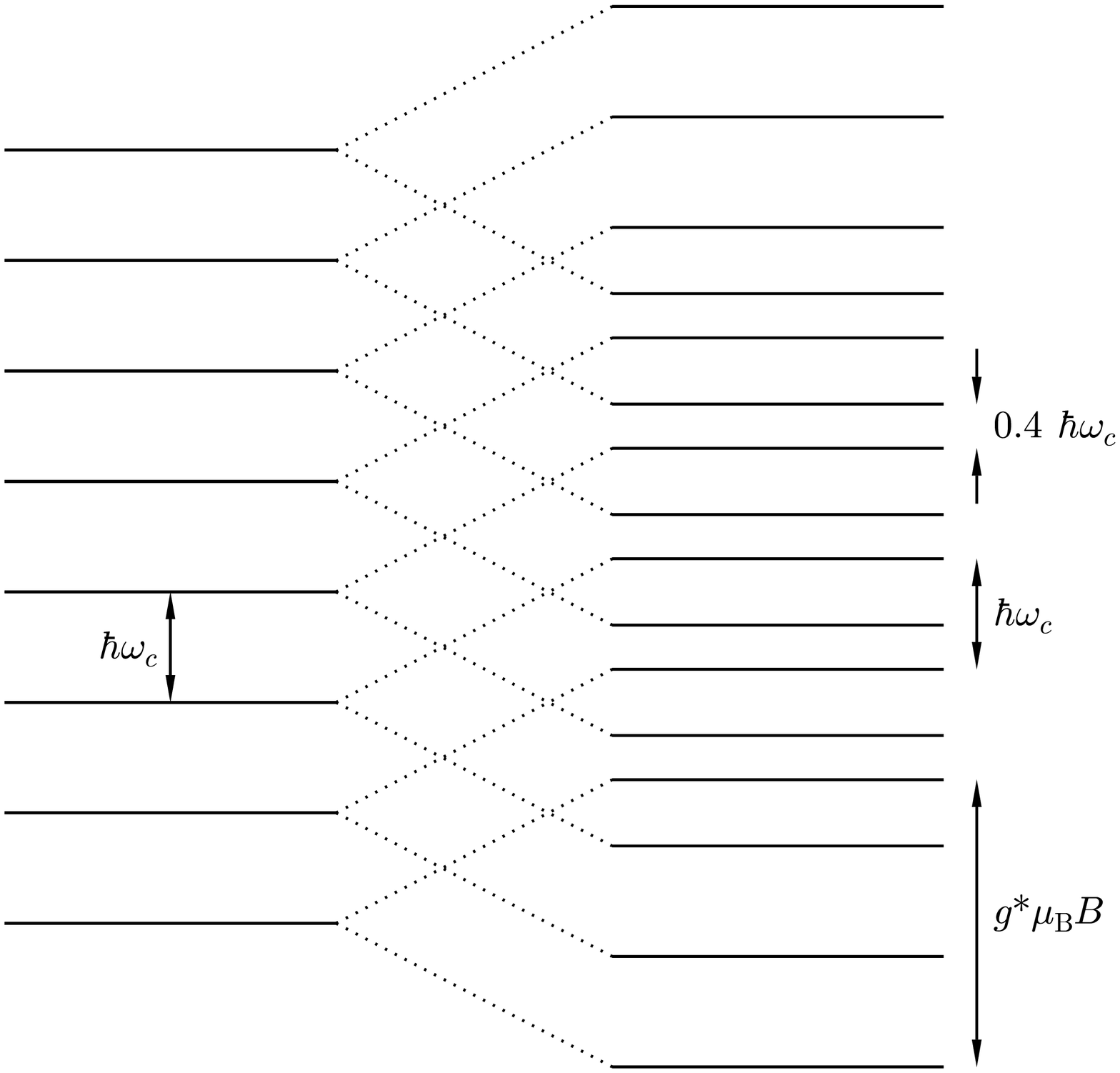}
\caption{The energy level spectrum of \cuscn~at $\theta=0$\deg~as deduced from
$g^*\mu^*_\alpha\approx5.2$, measured at $\sim0.5$~K in the
low-field region. There are two splittings, one of
$\hbar\omega_c$, which results in the SdH oscillations of the
fundamental frequency, and one of $0.4~\hbar\omega_c$, which at
sufficiently low temperatures and high fields will result in the
observation of harmonics of the fundamental frequency.}
\label{cuscnsplit}
\end{figure}
}

\def \Q1dsim{
\begin{figure}[tbp]
\centering \includegraphics[height=12cm]{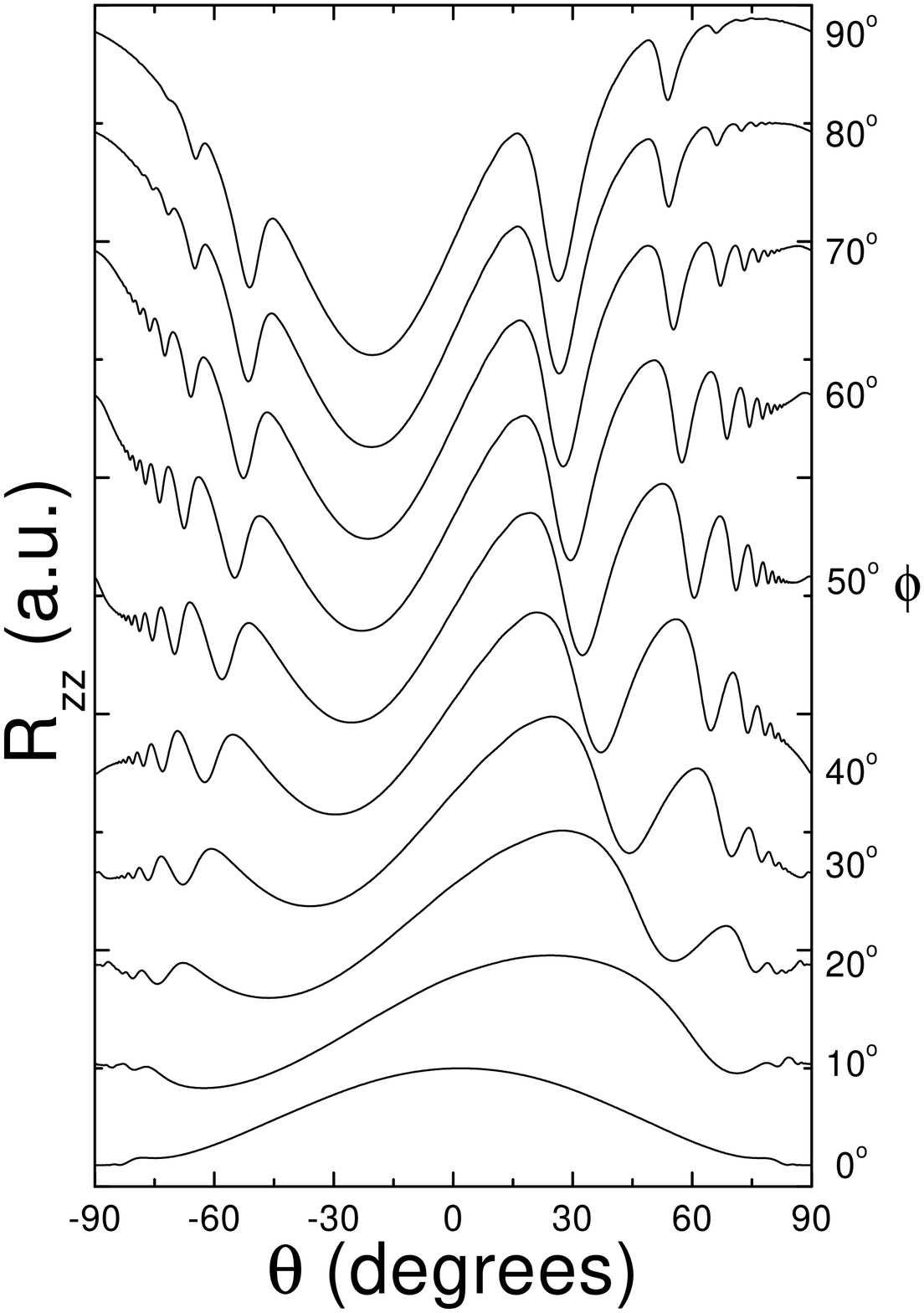}
\caption{The simulated interplane resistance resulting from
solving the Chambers formula numerically for the Q1D \fs~sheets of
\cuscn, as described by \Eref{cuscndisp}. The $\theta$-dependences
are shown for a fixed magnetic field of 42~T and a selection of values of the azimuthal angle,
$\phi$. The Lebed magic angle effect dominates the
magnetoresistance, except at low $\phi$-angles where the
Danner-Kang-Chaikin oscillations are seen around $\theta=90$\deg.}
\label{q1dsim}
\end{figure}
}

\def \q1dsimres{
\begin{figure}[tbp]
\centering \includegraphics[height=6cm]{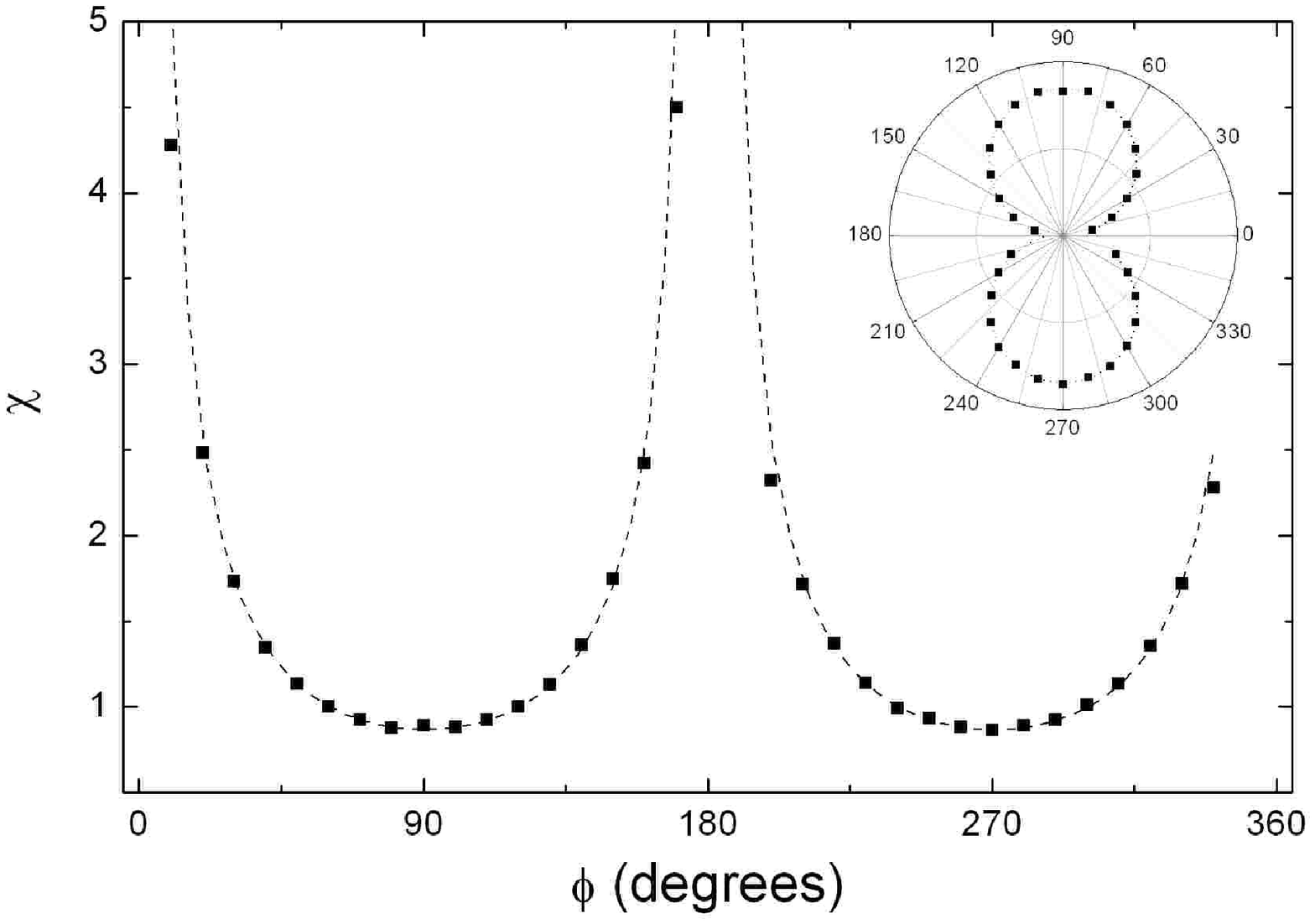}
\caption{The $\phi$-dependence of $\chi$, deduced from the frequency of the
simulated Lebed magic angle dips. The dotted line is a fit to
\Eref{lebedphidep}. The insert shows the polar plot of
$k_\parallel^{\rm max}$ versus $\phi$ that would result if the Q1D
features were mistaken for Q2D Yamaji oscillations, with the
dotted line representing a fit to \Eref{kpara}.} \label{q1dsimres}
\end{figure}
}

\def \aQ2dsim{
\begin{figure}[tbp]
\centering \includegraphics[height=12cm]{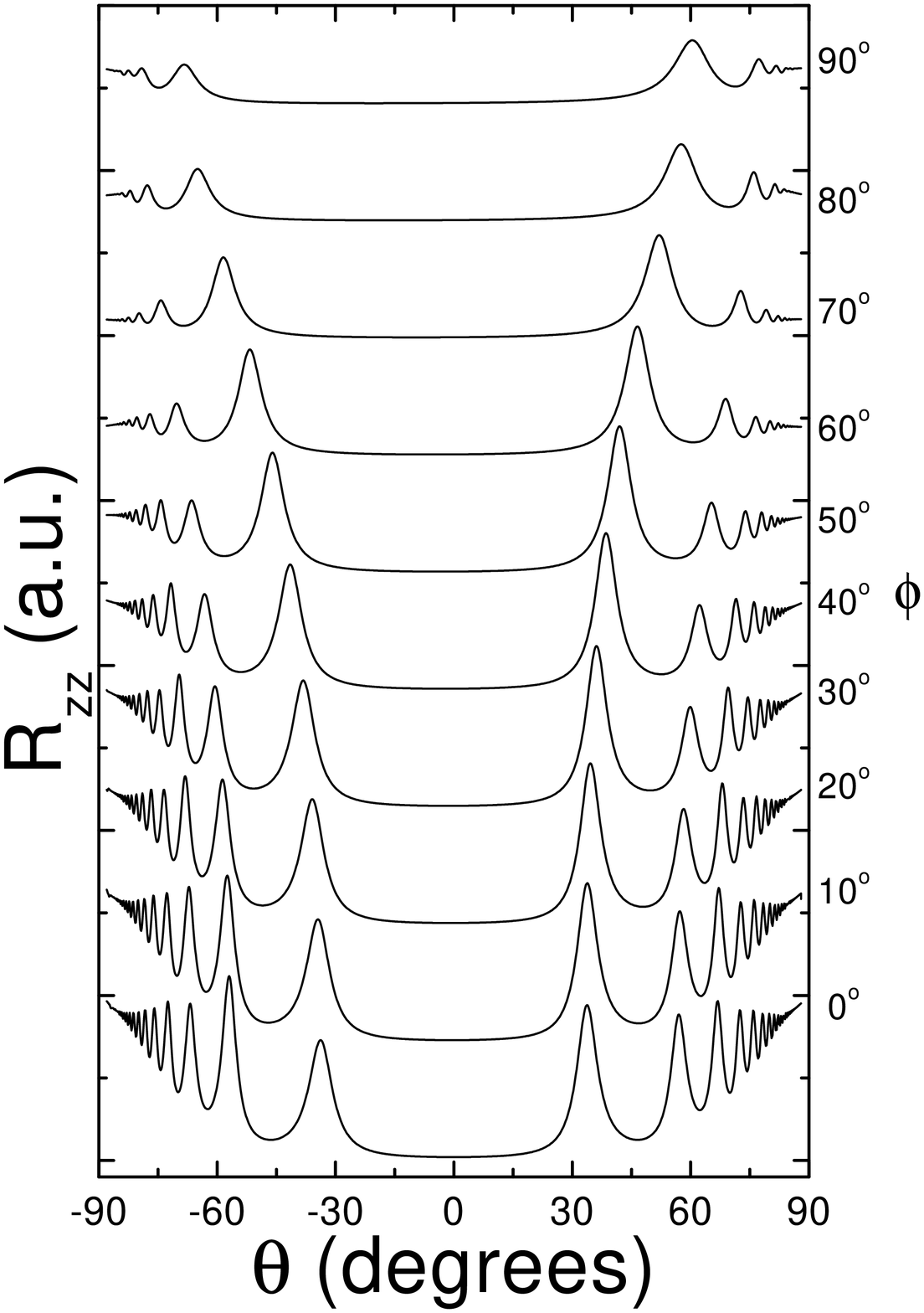}
\caption{The simulated interplane resistance resulting from
solving the Chambers formula numerically for the Q2D \fs~pockets
of \cuscn, as described by \Eref{cuscndisp}. The
$\theta$-dependences are shown for a fixed magnetic field of 42~T and a selection of values of the
azimuthal angle, $\phi$. The Yamaji oscillations dominate the
magnetoresistance, taking their maximum frequency when the
in-plane field is applied along $\phi=0$\deg, which is parallel to
the major axis of the Q2D pocket.} \label{q2dsim}
\end{figure}
}

\def \aq2dsimres{
\begin{figure}[tbp]
\centering \includegraphics[height=6cm]{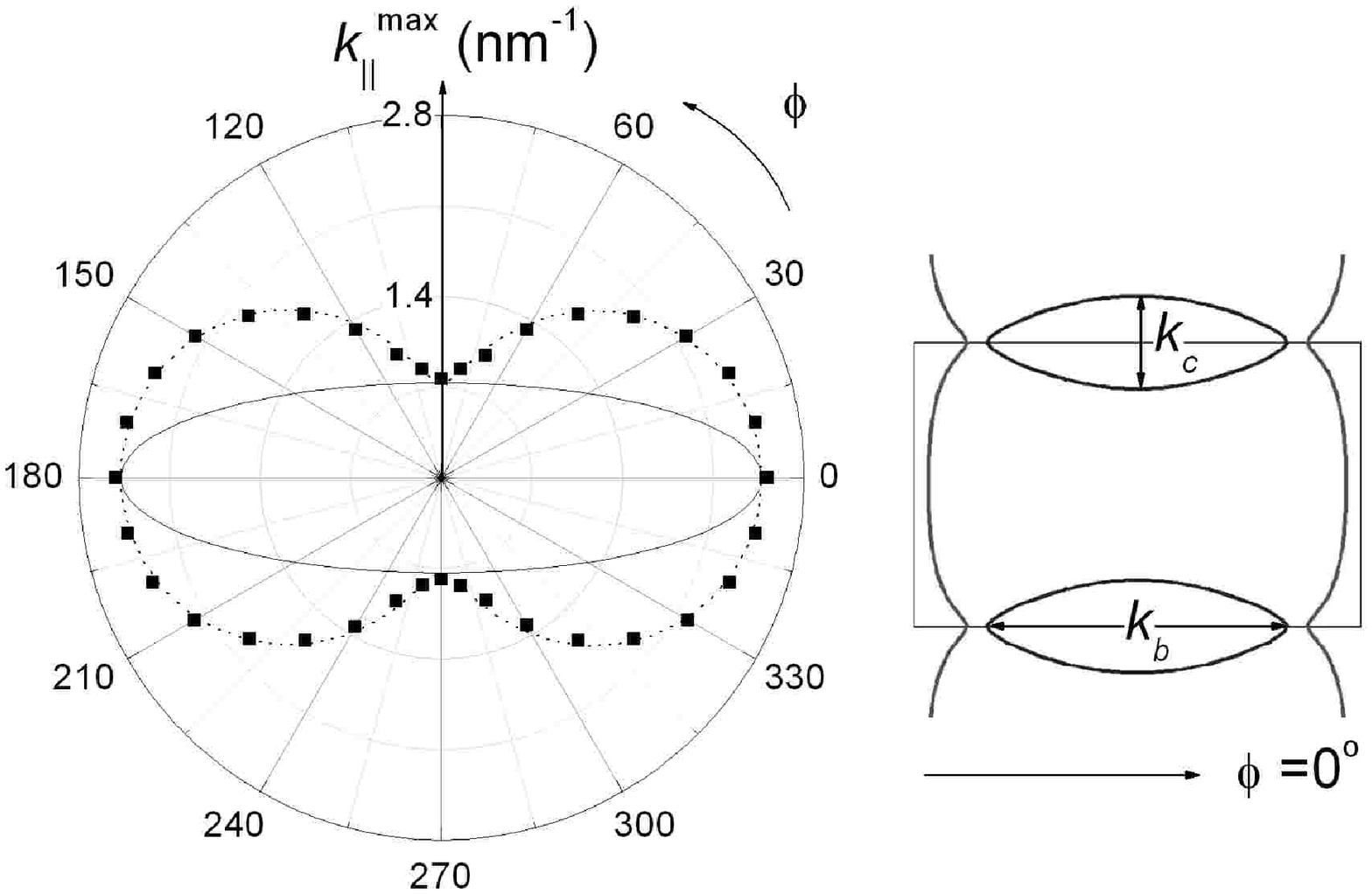}
\caption{Left: The data points are the $k_\parallel^{\rm max}$
values as obtained from the frequency of the Q2D Yamaji
oscillations in the simulated resistance, the dotted line is a fit
to \Eref{kpara}, and the solid line is the resulting \fs~pocket.
Right: A reminder of the in-plane \fs~of \cuscn.}
\label{aq2dsimres}
\end{figure}
}

\def \sasamro{
\begin{figure}[tbp]
\centering \includegraphics[height=12cm]{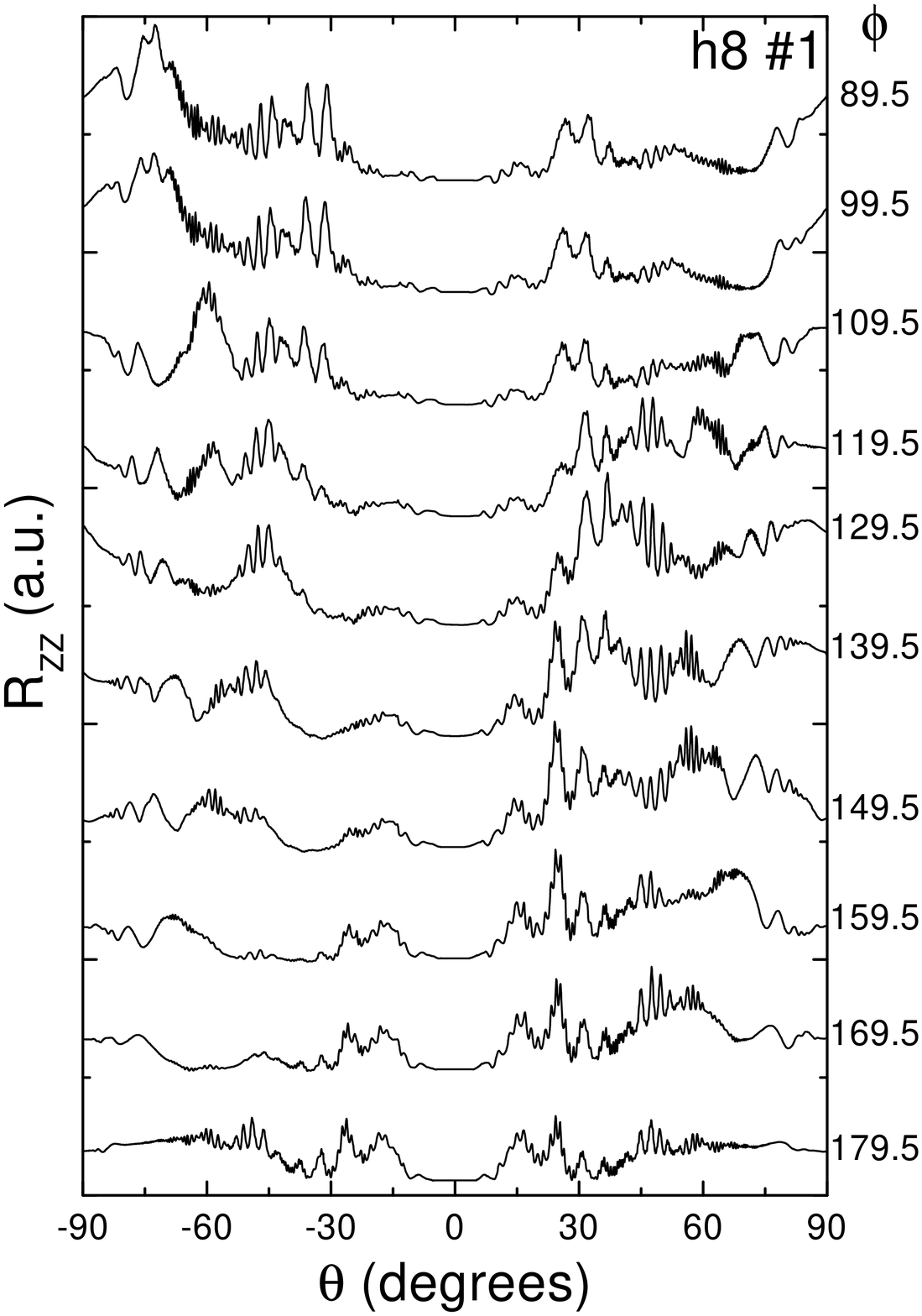}
\caption{The angle-dependent interlayer magnetoresistance of $h8$
\cuscn~at various values of the azimuthal angle, $\phi$. $T\approx500$~mK and
$B=42$~T.} \label{sasamro}
\end{figure}
}

\def \ah8results{
\begin{figure*}[tbp]
\centering \includegraphics[height=7cm]{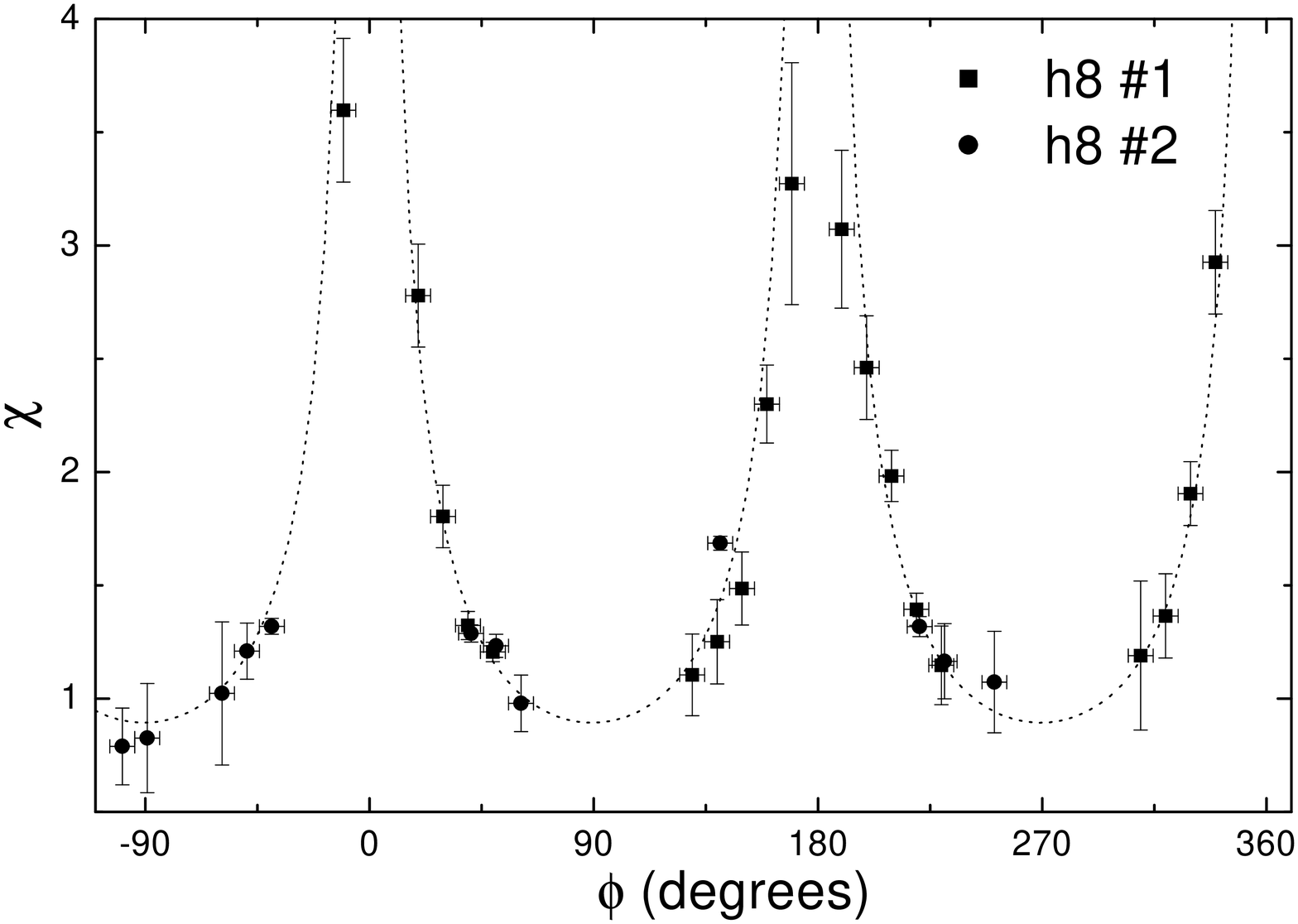}
\includegraphics[height=7cm]{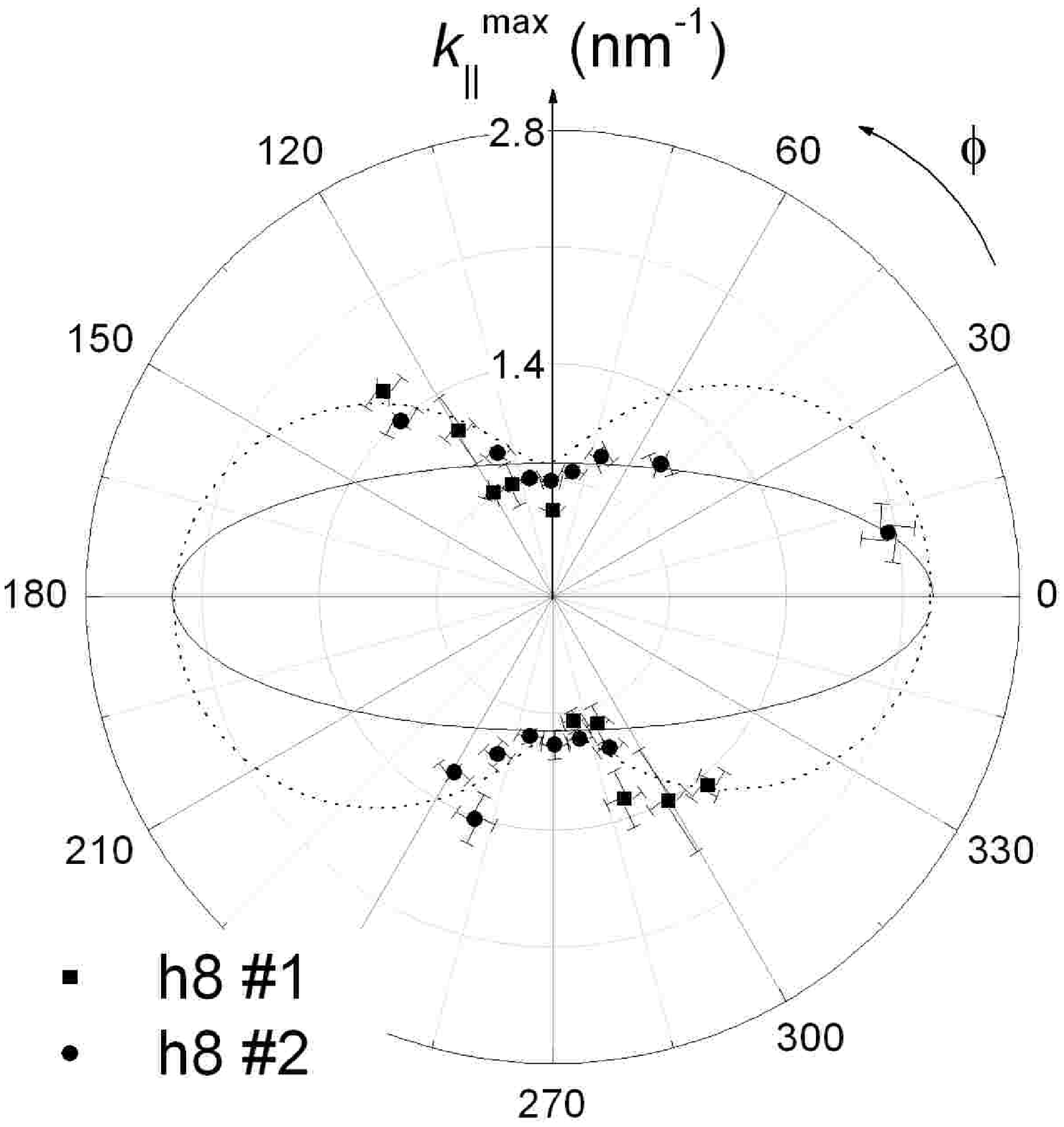}
\caption{Left: The value of $\chi$, obtained from the frequency of the resistance dips caused by the Lebed magic angle effect, at various values of the calibrated
azimuthal angle, $\phi$. The dotted line is a fit to \Eref{lebedphidep}. Right: The values of $k_\parallel^{\rm max}(\phi)$ obtained from an Yamaji analysis of the peaks in resistance at various values of the calibrated azimuthal angle, $\phi$. The dotted line is a fit to \Eref{kpara} constrained so
that the resulting Q2D pocket (solid line) has an area corresponding to the measured fundamental frequency. In both figures the squares are the data from $h8$ sample \#1, the circles are the data from $h8$ sample \#2} \label{ah8res}
\end{figure*}
}

\def \ch8resall{
\begin{figure}[tbp]
\centering \includegraphics[height=6cm]{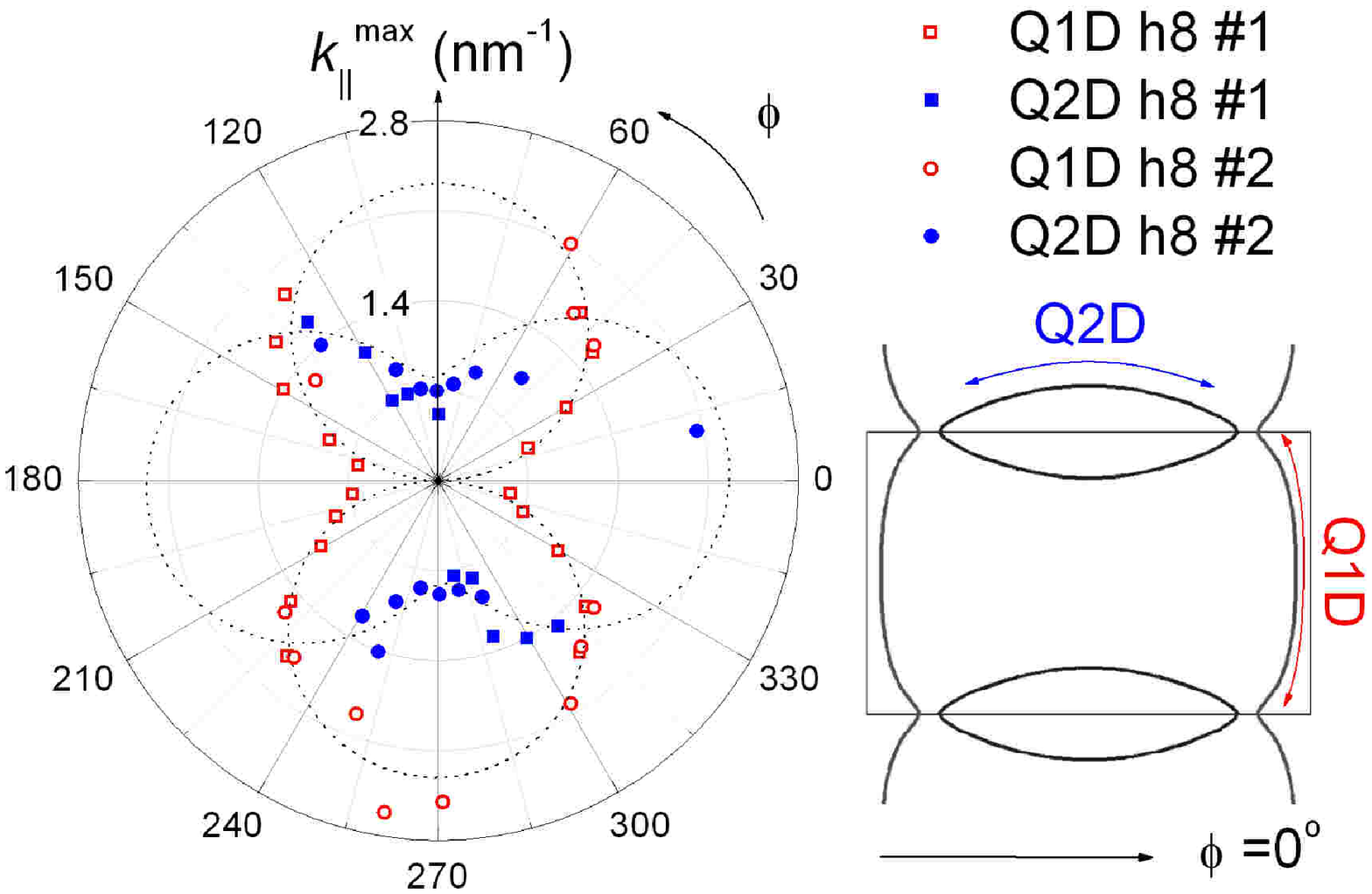}
\caption{The result of analysing all the AMRO as if they were
Yamaji oscillations. The squares are the data from $h8$
sample \#1, the circles are the data from $h8$ sample \#2,
the hollow symbols are data from Lebed magic angle dips, the solid
symbols are data from Yamaji peaks, and the dotted lines are the
fitted curves from Figure~{\ref{ah8res}}.
This shows that the Q1D effects dominate when the field is nearly
perpendicular to the sheets, and the Q2D effects dominate when the
field is nearly perpendicular to the flatter edge of the
\fs~pockets.} \label{h8resall}
\end{figure}
}

\def \3rdang{
\begin{figure}[tbp] \vspace{0.5cm}
\centering \includegraphics[height=8cm]{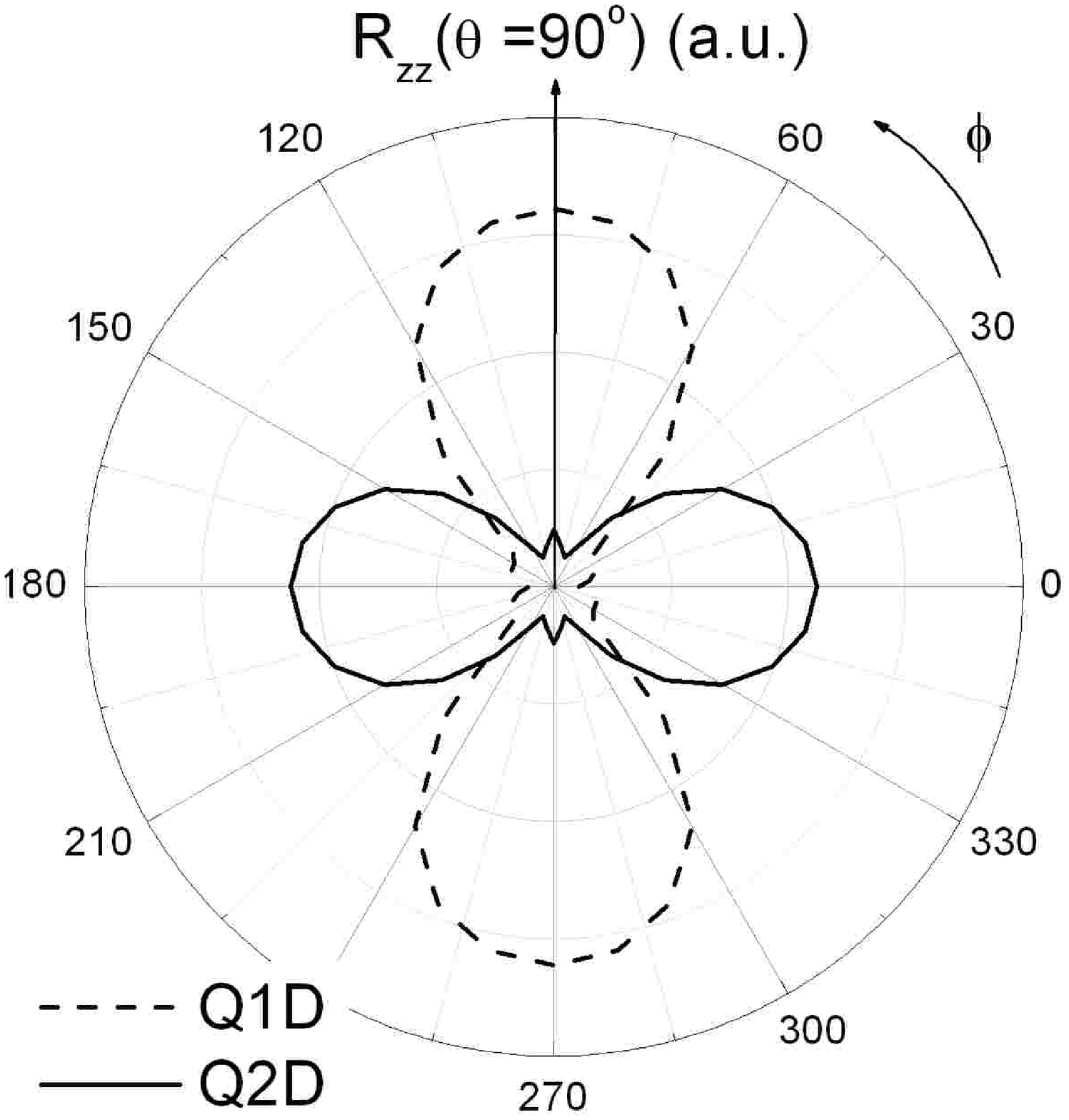}
\caption{$\phi$-dependence of the simulated interlayer resistance
at $\theta=90$\deg~for both the Q1D and Q2D sections of \fs~at a fixed field of 42~T. This
is the same geometry as the third angular effect.} \label{3rdang}
\end{figure}
}

\def \bd8results{
\begin{figure*}[tbp] \vspace{0.8cm}
\centering \includegraphics[height=7cm]{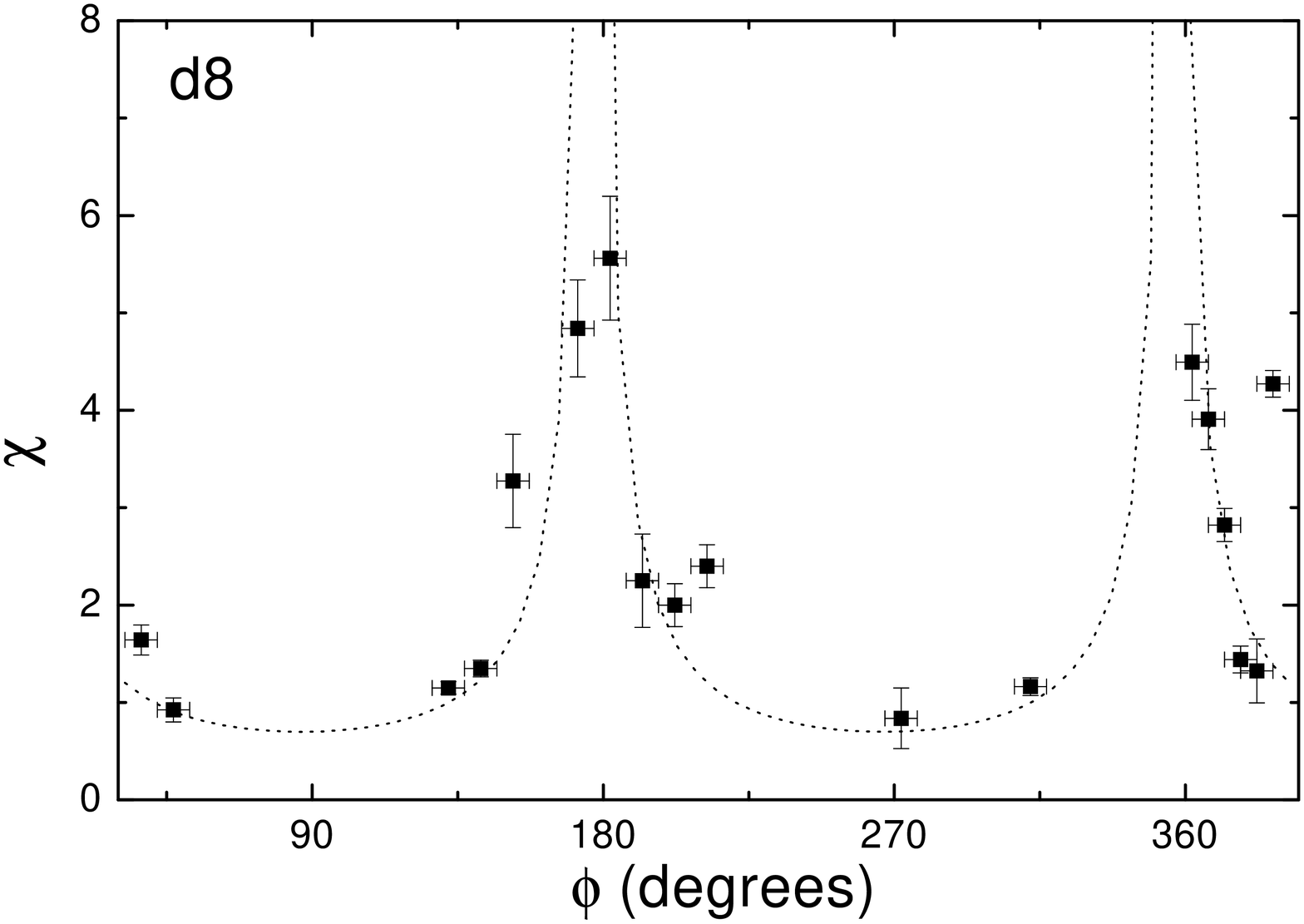}
\includegraphics[height=7cm]{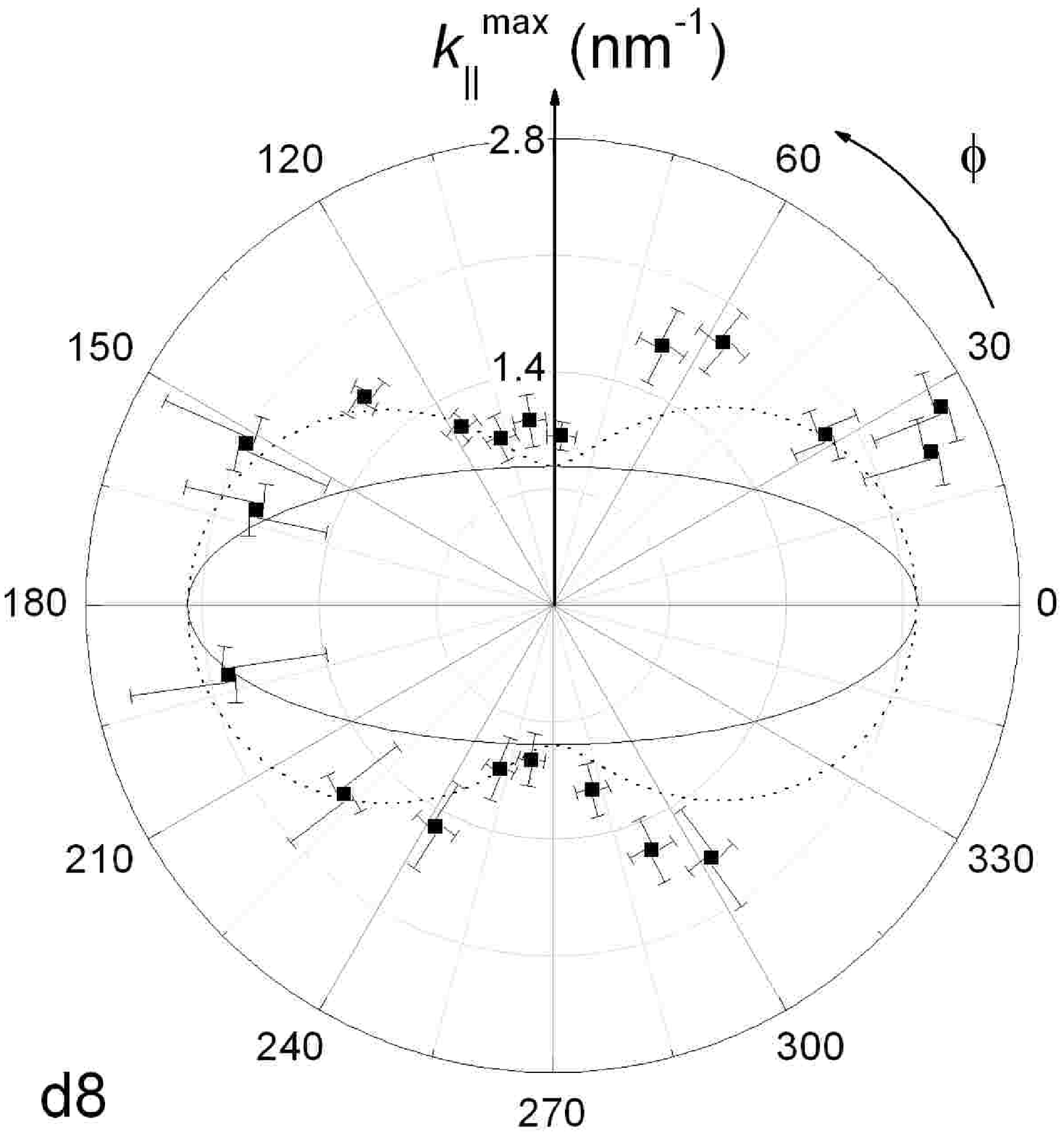}
\caption{Results for the $d8$ sample. Left: The value of $\chi$, obtained from the frequency of the resistance dips caused by the Lebed magic angle effect, at various values of the calibrated azimuthal angle, $\phi$. The dotted line is a fit to \Eref{lebedphidep}. Right: The values of $k_\parallel^{\rm max}(\phi)$ obtained from an Yamaji analysis of the peaks in resistance at various values of the calibrated azimuthal angle, $\phi$. The dotted line is a fit to \Eref{kpara} constrained so
that the resulting Q2D pocket (solid line) has an area corresponding to the
measured fundamental frequency.} \label{bd8res}
\end{figure*}
}

\def \squitorbs{
\begin{figure}[tbp]
\centering
\includegraphics[angle=90, height=3.5cm]{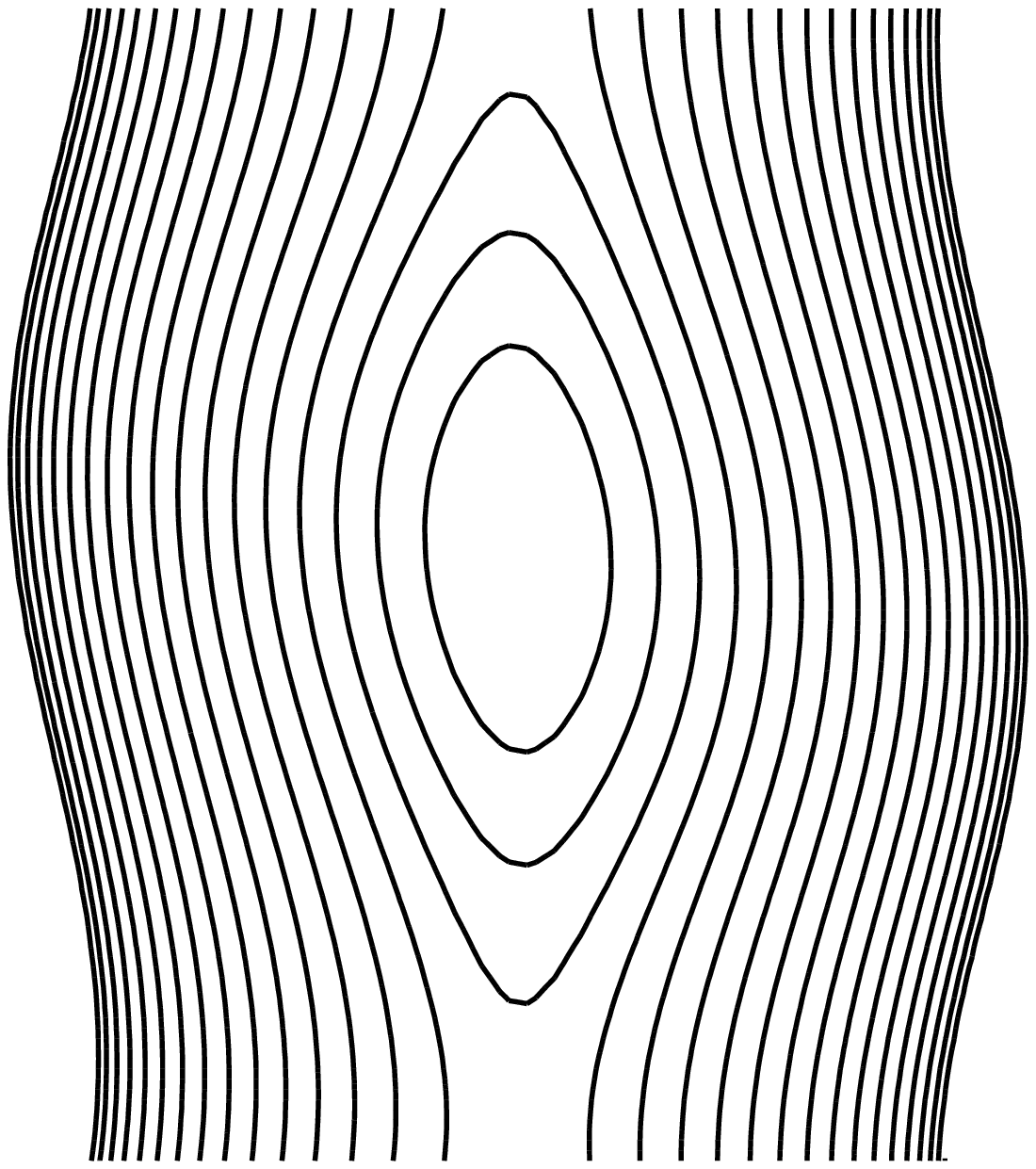}
\hspace{1cm}
\includegraphics[angle=90, height=3.5cm]{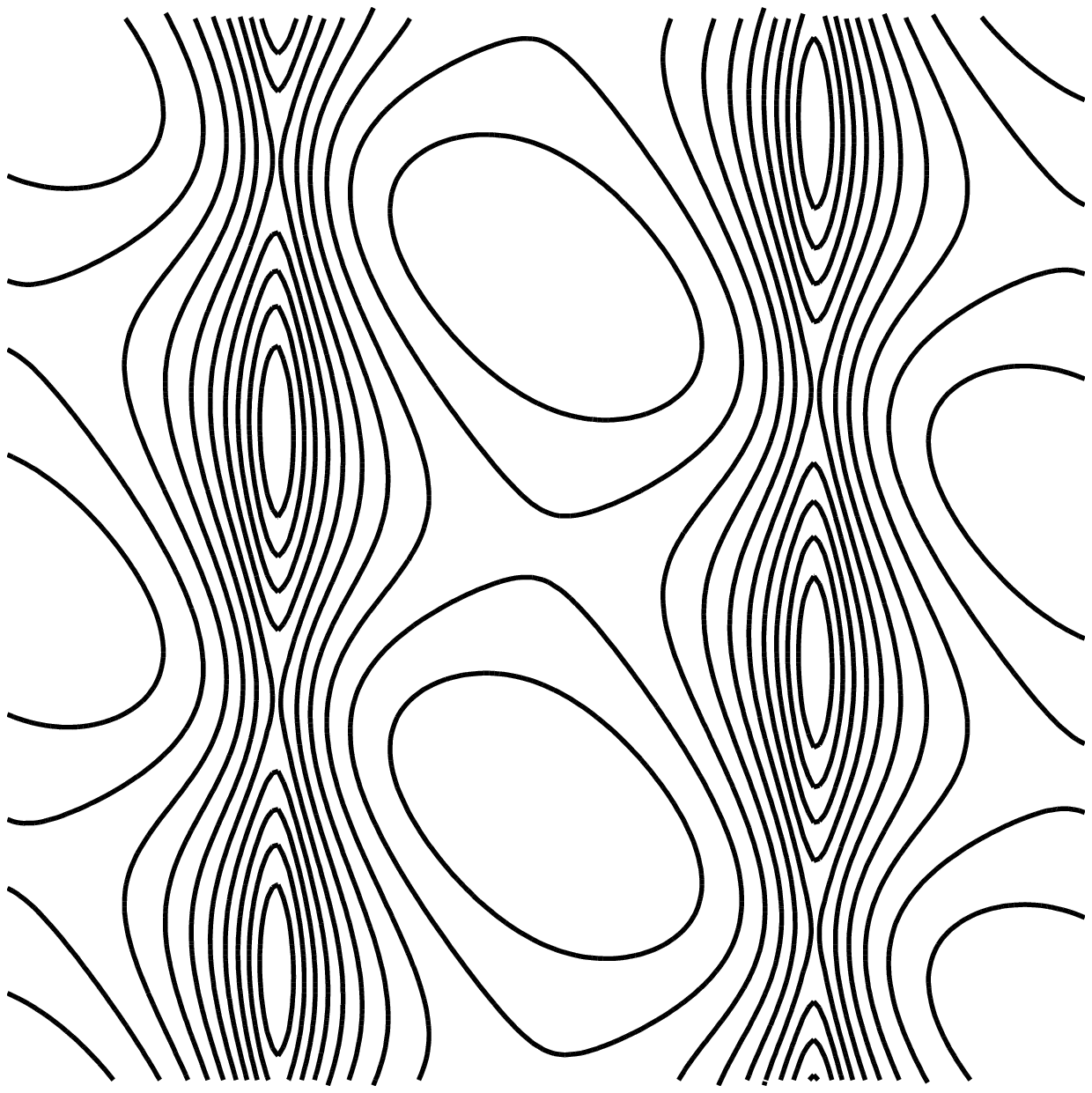}
\caption{Examples of the orbits possible on the Q2D (left) and Q1D
(right) sections of the \fs~defined by \Eref{cuscndisp} when the
magnetic field is applied parallel to the highly conducting
planes. All the orbits tend to average the interlayer velocity to
zero, and hence produce an increase in the interlayer resistance.
For the purposes of the illustration the interlayer transfer
integral has been exaggerated compared to its experimentally
determined value.} \label{squitorbs}
\end{figure}
}

\def \squitwidth{
\begin{figure*}[tbp]
\centering \includegraphics[height=6.1cm]{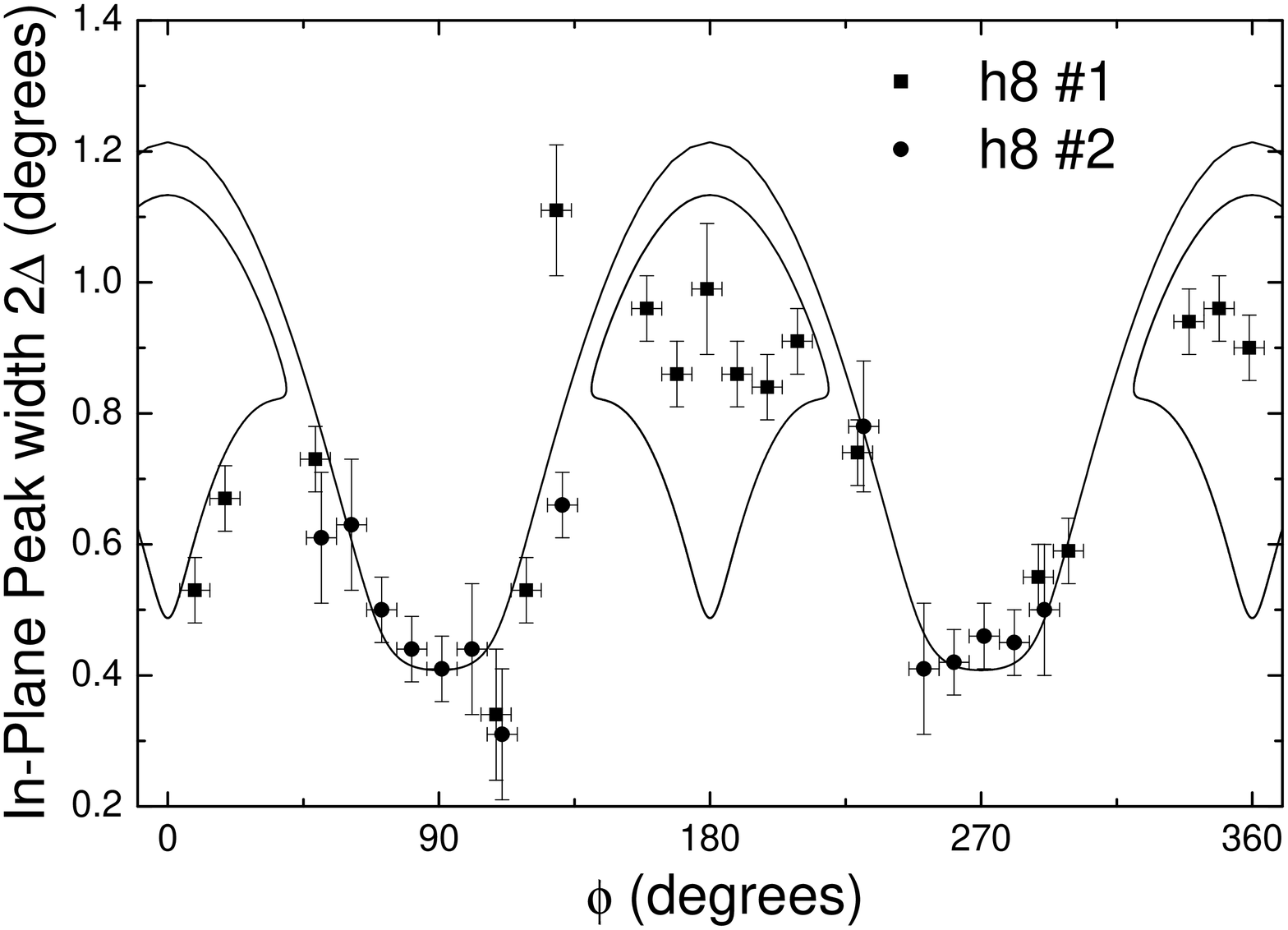}
\centering \includegraphics[height=5.9cm]{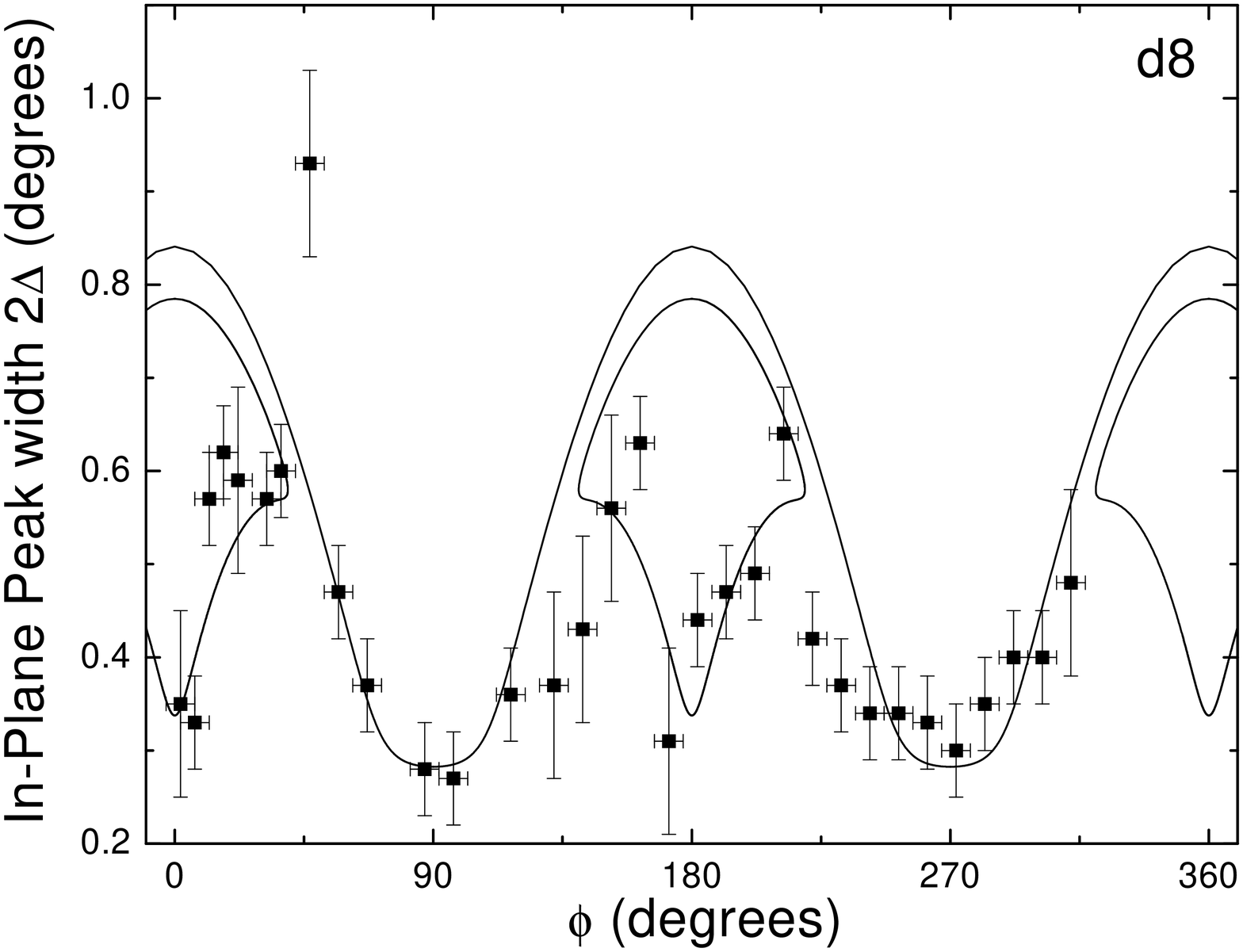}
\caption{Azimuthal angle dependence of the angular width,
2$\Delta$, of the in-plane peak for $h8$ (left) and $d8$ (right) \cuscn~at $B=42$~T,
$T\approx500$~mK. The points are the experimentally measured
widths and the solid lines are the simulated width with $t_a(h8)=0.065\pm0.007$~meV and
$t_a(d8)=0.045\pm0.005$~meV. The continuous curve represents the contribution
from the Q2D \fs~pockets, whereas the closed curves are those from
the Q1D sheets.} \label{squitwidth}
\end{figure*}
}

\newcommand{\chemlett}{{\it Chem.~Lett.\ }}

\newcommand{\jap}{{\it J.~Appl.~Phys.\ }}

\newcommand{\jdpI}{{\it J.~Physique I}}

\newcommand{\jpcm}{{\it J.~Phys.:~Condens.~Matter }}

\newcommand{\jpsj}{{\it J.~Phys.~Soc.~Japan }}

\newcommand{\phyc}{{\it Physica~C }}

\newcommand{\pps}{{\it Proc.~Phys.~Soc.\ }}

\newcommand{\rpp}{{\it Rep.~Prog.~Phys.\ }}

\newcommand{\ssc}{{\it Sol.~Stat.~Commun.\ }}

\newcommand{\sm}{{\it Synth.~Met.\ }}

\begin{document}

\preprint{APS/CuSCN}

\title{Angle Dependent Magnetoresistance of the Layered Organic
Superconductor \cuscn : \ Simulation and Experiment}

\author{P. A. Goddard$^1$} \email{pgoddard@lanl.gov} \author{S. J.
Blundell$^2$} \author{J. Singleton$^1$} \author{R. D. McDonald$^1$} 
\author{A.~Ardavan$^2$} \author{A. Narduzzo$^2$} \author{J. A. Schlueter$^3$} \author{A. M. Kini$^3$} \author{T. Sasaki$^4$}
\affiliation{$^1$National High Magnetic Field Laboratory, Los Alamos National Laboratory, TA-35, MS-E536, Los Alamos, New Mexico 87544, USA}
\affiliation{$^2$Department of Physics, University of Oxford, Clarendon Laboratory, Parks Road, Oxford, OX1 3PU, United Kingdom}
\affiliation{$^3$Materials Science Division, Argonne National Laboratory, Argonne, Illinois 60439, USA}
\affiliation{$^4$Institute for Materials Research, Tohoku University, Sendai 980-77, Japan}

\date{\today}

\begin{abstract}  

The angle-dependences of the magnetoresistance of two different isotopic substitutions (deuterated and undeuterated) of the layered organic superconductor \cuscn~are presented. The angle dependent magnetoresistance oscillations (AMRO) arising from the quasi-one-dimensional (Q1D) and quasi-two-dimensional (Q2D) \fs s in this material are often confused. By using the Boltzman transport equation extensive simulations of the AMRO are made that reveal the subtle differences between the different species of oscillation. No significant differences are observed in the electronic parameters derived from quantum oscillations and AMRO for the two isotopic substitutions. The interlayer transfer integrals are determined for both isotopic substitutions and a slight difference is observed which may account for the negative isotope effect previously reported~\cite{schlueter01}. The success of the semi-classical simulations suggests that non-Fermi liquid effects are not required to explain the interlayer-transport in this system.

\end{abstract}

\maketitle

\section{\label{intro}Introduction} \begin{sloppypar}

\cuscn~is probably the most popular and best characterised material
out of all the organic charge transfer salts based on the ET molecule.
Its attraction to experimentalists lies in its exceedingly simple \fs,
which consists of two elliptical quasi-two-dimensional (Q2D) pockets and
a pair of warped quasi-one-dimensional (Q1D)
sheets~\cite{sasaki90,sasaki91}~(see \Fref{cuscnfs}). The prospect
of understanding the complex transport properties of the organic salts
seems more within reach for this material than for others that show
similar behaviour but have more complicated \fs s.

Several theoretical models of the unconventional superconductivity observed in \cuscn~and related materials suggest that the superconducting pairing mechanism may be mediated by antiferromagnetic spin fluctuations~\cite{schmalian98,kondo98,kuroki99}. These models are found to be sensitive to the degree to which the \fs~of the material can nest; the higher the nestability the more likely this pairing is to be successful. Two-dimensional \fs s are clearly better able to nest than three-dimensional ones, and so tests of the dimensionality of \cuscn~also test these theoretical models. 

In this paper the low temperature angle-dependence of the
magnetoresistance in deuterated and undeuterated samples of \cuscn~is
studied in detail in magnetic fields significantly higher than in-plane
upper critical field. This is the first time that comprehensive
measurements like these have been made. Their purpose is to completely
determine the parameters that define the transport in this material, to
locate any differences between these parameters for the two isotopic
substitutions that might shed light on the disparity between their
superconducting critical temperatures~\cite{schlueter01}, and to address
the question of whether it is possible to describe all aspects of the
normal state transport within the bounds of Fermi liquid theory.

As with all the organic conductors in this class, the ET molecules form
the highly conducting layers, separated by layers of the anion, with the
long axis of the ET molecule at a small angle to the interlayer
direction. In the $\kappa$-phase salts the ET molecules associate into
pairs, or dimers, each of which collectively donates one electron to the
anions, leaving behind a mobile hole~\cite{urayama88}. There are two
dimers, and thus two holes per unit cell, and so, because the dispersion
is nearly isotropic in the $bc$-plane, this leads to a roughly circular
\fs~which has the same area as the first \bz~\cite{sasaki90}. The
\bz~itself reflects the rectangular cross-section of the unit cell and
the \fs~cuts the \bz~boundaries on its long side. At these points a gap
opens up which splits the \fs~into the Q1D and Q2D
sections~\cite{sasaki90,sasaki91}. The result is shown in the top part
of \Fref{cuscnfs}.

\cuscnfs

The shape of the \fs~in the $k_xk_y$-plane has been confirmed by the observation of magnetic quantum oscillations~\cite{sasaki90,caulfield94}. The frequency of the
quasiparticle orbits about the circumference of the Q2D pockets
($\alpha$-orbits) is found to be 600~T which corresponds to about 15~\%
of $A_{BZ}$, the area of the cross-section of the first \bz~in the
$k_x$$k_y$ plane~\cite{caulfield94}. Above 20~T magnetic breakdown is
observed as it becomes possible for some of the quasiparticles to
bridge the energy gap between the two \fs s and make the large $\beta$-orbit, whose frequency is found to be 3920~T~\cite{caulfield94}. This corresponds to an area equal to $A_{BZ}$ to within a few percent, as would be expected from the
considerations discussed above.

\cuscn~has a monoclinic structure and the transfer integral between the
layers lies parallel to the $a$ lattice parameter which is inclined at
angle of 110.3\deg~to the highly conducting
$bc$-planes~\cite{urayama88}. This transfer integral, $t_a$, is much
smaller than those within the planes and results in a slight warping of
the \fs~perpendicular to the direction of $t_a$ in $k$-space. This is
shown in the lower portion of \Fref{cuscnfs}. The validity of this
picture of the \fs~as a three-dimensional object is discussed in Reference~\cite{squit}.

\section{Overview of the features observed in magnetotransport}

\Fref{typicalsweep} shows two typical $\theta$-dependences (where $\theta$~is the angle between the magnetic field and the normal to the conducting layers) of the interlayer magnetoresistance of \cuscn~in fixed magnetic fields of 27~T and 42~T and at an azimuthal angle of 149\deg.  In such high fields a whole host of features are observed in an interlayer transport measurement of \cuscn, e.g.\ Shubnikov de Haas (SdH) oscillations,  magnetic breakdown, and Q1D and Q2D angle dependent magnetoresistance oscillations (AMRO). This means that for a typical $\theta$-rotation  the magnetoresistance is rich in features as \Fref{typicalsweep} illustrates.

\typicalsweep

The upper plot shows the data taken at 42~T and the lower at 27~T. In the upper plot the field perpendicular to the layers around
$\theta=0$\deg~is sufficient for the effects of magnetic breakdown to be
observed and the fast SdH oscillations due to the $\beta$-orbit are
clearly seen. The slower oscillations due to the $\alpha$-orbit are seen
in both plots and persist to higher angles. The amplitudes of these
oscillations are modulated and they disappear at certain
$\theta$-angles, these nodes are known as {\it spin-zeroes} and are
caused by Zeeman splitting of the Landau levels~\cite{nicholas88}. An
analysis of this effect is dealt with in~\Sref{g}.

The positions of the features at $\theta$-angles greater than about
$\pm$70\deg~are seen to be independent of the magnitude of the magnetic
field, which reveals them to be AMRO of one variety or another. Four different types of AMRO are possible in the interlayer resistivity (\rhozz) of \cuscn. These are the Danner-Kang-Chaikin oscillations~\cite{dkc}, the third angle effect~\cite{osada96} and the Lebed magic angle effect~\cite{blundell96} which all arise from orbits on the Q1D \fs~section, and the Yamaji oscillations arising from orbits on the Q2D \fs~section~\cite{peschansky91, kartsovnik92}. In the semi-classical picture all the AMRO are caused by the degree to which the velocity components of the quasiparticle are averaged over the series of orbits that appear at a certain inclination angle. In particular, the orbits that are possible in the region of the Yamaji angles are very successful in averaging the interlayer velocity towards zero, thus peaks are seen in the interlayer resistance~\cite{peschansky91, kartsovnik92}. In contrast, the orbits that occur at the Lebed magic angles are not as successful at averaging the interlayer velocity towards zero as those possible at the other angles and so dips in \rhozz~are observed~\cite{blundell96}. There are other theories that can explain the effects observed at the Lebed magic angles. Lebed's own argument describes electron-electron correlations whose magnitudes change when the field is directed along the magic
angles~\cite{lebed95}. Another theory has regions of $k$-space
where the scattering rate takes a large value (\fs~hotspots)
accounting for the AMRO~\cite{hotspots}. However, such theories
are complicated and need only be invoked when the semiclassical
approach fails to account for the experimental
observations. It will be shown by the simulations described in \Sref{amro} that the semi-classical explanation is sufficient in the case of \cuscn.

In the upper plot of~\Fref{typicalsweep} a small peak
is observed when the field lies very close to the in-plane direction,
$\theta\approx90$\deg. This is the in-plane peak feature mentioned in Reference~\cite{squit}. It will be discussed further in \Sref{squit}.
Around $\theta=90$\deg~in the lower plot the in-plane peak is obscured by the
large dip that indicates the onset of a superconducting transition. This
occurs because there is a considerable anisotropy in the upper critical
field of this material, and a field of 27~T is not sufficient to
suppress the superconducting state when applied in a nearly in-plane
direction~\cite{loff}.

It should also be noted from \Fref{typicalsweep} that the amplitude of
all the features in the magnetoresistance increase with increasing field, and that the plots are not symmetrical about $\theta=0$\deg, reflecting the monoclinic symmetry of the crystal structure.

\section{Parameterising the \fs} \label{fs}

It has been shown that the measured intralayer \fs~of \cuscn~can be
reproduced using a dispersion relation derived from a tight binding
model using the ET dimer as its base
unit~\cite{schmalian98,squit,caulfield94}. In this way the intradimer
transfer integral, $t_d$, can be ignored and shape of the \fs~depends
upon interdimer transfer integrals $t_b$, $t_{c1}$, $t_{c2}$ and the
Fermi energy, \Ef. The dispersion found in this manner is known as the
{\it effective dimer model} and is given by 

\begin{eqnarray} E({\bf k})&=&2t_b\cos(k_xb) \nonumber \\ 
&& \pm \cos\left(\frac{k_xb}{2}\right)\sqrt{t_{c1}^2+t_{c2}^2+2t_{c1}t_{c2}\cos (k_yc)} \mbox{~~}\label{effectivedimer} 
\end{eqnarray} 
where the + and $-$ signs result in the Q1D and the Q2D sections of the
\fs~respectively~\cite{schmalian98,caulfield94}.

The effective dimer model is used in the semi-classical calculations of
Sections~{\ref{amro}} and {\ref{squit}}. These do not take
account of quantum effects such as Shubnikov de Haas oscillations or
magnetic breakdown, and are assumed to be in the ``low-field'' region
where breakdown does not occur. It will be seen that this is a
reasonable assumption in both cases. This means that the effect of the
energy gap, i.e.\ the difference between $t_{c1}$ and $t_{c2}$, can at
first be neglected and the shape of the in-plane \fs~depends only upon
the ratios $E_{\rm F}/t_c$ and $t_b/t_c$, where $t_c$ is an average of
$t_{c1}$ and $t_{c2}$~\cite{schmalian98,squit}.

It is possible to obtain values for these ratios by adjusting them to
reproduce the areas of the $\alpha$ and $\beta$ \fs~orbits. Once this
is done $t_b$, $t_c$ and \Ef~can be uniquely specified by fitting to the
effective mass of the $\beta$-orbit as found from SdH
oscillations, using the expression $\delta
A_\beta/\delta E=2\pi m_\beta^*/\hbar^2$~\cite{shoenberg}. Note that it also possible to fit to the mass of the $\alpha$-orbit to obtain slightly different results. However, as the masses are derived from quantum oscillations, they are orbitally
averaged, and so the $\alpha$-mass will be dominated by the extremely
pointed regions of the Q2D pockets. The breakdown orbit does not have these pointed regions and thus it is the $\beta$-mass that
is used in the fitting procedure.

The energy gap can now be reintroduced in order to specify $t_{c1}$ and
$t_{c2}$. In the region of the gap, $\cos(k_cc)=-1$ and
$\cos(k_bb/2)\approx0.5$, so that \mbox{$E_g\approx2(t_{c1}-t_{c2})$}.
From the magnetic breakdown, $E_g$ is estimated to be
7.8~meV~\cite{harrison96mb,squit}. Any inaccuracies in this value will
not lead to errors in the size of the \fs~produced by
\Eref{effectivedimer}, but could lead to small discrepancies in the
exact dimensions of the $\alpha$-pocket.

The \fs~parameters obtained in this manner are as follows:
\mbox{$t_b=14.87$~meV}; \mbox{$t_{c1}=26.65$~meV};
\mbox{$t_{c2}=22.75$~meV}; and \mbox{$E_{\rm F}=-19.12$~meV}. Note that
these values of $t$ are effective transfer integrals, and incorporate
the effects of electron-phonon and electron-electron interactions as
they are derived from magnetic quantum oscillation data (see~\cite{harrison99}
for a discussion). Note also that \Ef~is taken relative to the zero
energy of the effective dimer model and not the bottom of the band. It
should therefore {\it not} be quoted as the Fermi energy of \cuscn.

\section{Experimental details} \label{cuscn:expdetails}

Four single crystal samples of \cuscn~were used in this study, made
using an electrocrystallisation method~\cite[and references therein]
{schlueter01}. All of the samples are black platelets of the order of
$0.7\times0.5\times0.1$~mm$^3$, with the plane of the plate
corresponding to the highly conducting layers. In one of the samples the eight terminal hydrogens of the ET molecules were
substituted by deuterium. In what follows the deuterated crystal will be
referred to as $d8$, and the hydrogenated crystals as $h8$.

The magnetoresistance measurements were made using standard 4-wire A.C.
techniques ($f=50-180$~Hz) with the current applied in the interplane
direction ($I=1-20~\mu$A).

All the samples were mounted on a two-axis rotator in a $^3$He cryostat.
In this rotator it is possible to continuously change the
$\theta$-angle, the angle between the magnetic field and the highly
conducting $bc$-planes, and discretely change the plane of rotation,
described by the azimuthal angle, $\phi$. An angular calibration
technique similar to that described in Reference~\cite{goddardetcl} was used when
misalignments of the sample that occur during cooling were found to be
significant. Temperatures down to 0.5~K are readily accessible.

\section{Results and discussion} \label{results}

\subsection{\chapstyle The Shubnikov de Haas oscillations}
\label{cuscn:sdh} 

Using the results of a fast Fourier transform analysis of
several SdH measurements the fundamental frequencies were found to be
$F_\alpha(h8)=599\pm3$~T; $F_\beta(h8)=3860\pm6$~T;
$F_\alpha(d8)=598\pm3$~T; and $F_\beta(d8)=3871\pm10$~T, all of which are in reasonable agreement with previous results~\cite{caulfield94,tim}.

The $\alpha$-mass of $h8$ \cuscn~has previously been found to be
$m_\alpha^*(h8)=3.5\pm0.1~m_e$~\cite{caulfield94} using a
Lifshitz-Kosevich analysis of the temperature dependence of the SdH
amplitudes~\cite{shoenberg}. Using the same method the equivalent $d8$-mass is
$m_\alpha^*(d8)=3.6\pm0.1~me$.

The Lifshitz-Kosevich analysis can also be applied to the
field-dependence of the SdH amplitudes at a constant temperature to find
values for the scattering time and the breakdown field. Using the technique outlined in Reference~\cite{harrison96mb} (but
correcting the erroneous minus sign that prefixes the breakdown term in
that reference) the amplitudes, $A$, are fitted with the function
\begin{equation}
\ln\left[A^*\frac{\sinh(\gamma_jT/B)}{\gamma_jT/B}\right]=
\ln[A_0]-\frac{\gamma_jT_{{\rm D},j}}{B}
+\ln[p^{n_{1j}}q^{n_{2j}}],\label{mb} \end{equation} where
$\gamma_j=(2\pi^2\mu^*_jl_jk_{\rm B})/(\hbar e)$, $p^2=\exp(-B_{0j}/B)$
and $q^2=1-p^2$~\cite{shoenberg}. $T_{\rm D}$ is the Dingle temperature
and is proportional to the scattering rate, $\mu^*_j=m^*_j/m_{\rm e}$,
$m_{\rm e}$ is the mass of an electron, $l_j$ is the harmonic index of
the orbit, $n_{1j}$ is the number of magnetic breakthrough points on the
orbit, and $n_{2j}$ is the number of Bragg reflection
points~\cite{shoenberg}. Note that this is derived from the
two-dimensional form of the Lifshitz-Kosevich formula with the magnetic
field directed perpendicular to the highly conducting layers. The amplitudes are obtained from the fast Fourier transform
spectrum of the oscillating part of the resistance. The field window
over which the Fourier transform is performed specifies the value of $B$
in the above equation such that $B^{-1}=(B_1^{-1}+B_2^{-1})/2$, where
$B_1$ and $B_2$ are respectively the start and end points of the field
window~\cite{kamm78}.

The functional form of the $T_{\rm D}$ and $B_0$ terms in the
Lifshitz-Kosevich formula are similar and so to obtain a satisfactory
fit the amplitudes of the first and second harmonics of the
$\alpha$-frequency must be fitted simultaneously~\cite{harrison96mb}. In
this case $n_{1}(\alpha)=n_{1}(2\alpha)=0$, $n_{2}(\alpha)=2$ and
$n_{2}(2\alpha)=4$. Thus, at high fields these amplitudes are attenuated
as quasiparticles are able to tunnel across the gap to the
$\beta$-orbit. The fits are shown in \Fref{lkfits} and the values
obtained are $\tau(h8)=2.3\pm0.2$~ps; $\tau(d8)=2.4\pm0.2$~ps;
$B_0(h8)=58\pm9$~T; and $B_0(d8)=39\pm10$~T. The values of the
scattering time obtained from the high field fits are in  close
agreement with those obtained from a fit to the low-field data where the
effects of breakdown may be neglected.

\lkfits

Note that all the results for the $d8$ sample are the same as those for
the $h8$ to within the experimental errors. The large errors on the
values of the breakdown field and the discrepancies between these values
and those independently obtained from similar data
($B_0(h8)=41\pm7$~T~\cite{harrison96mb}) and even the same data
($B_0(h8)=41\pm5$~T and $B_0(d8)=30\pm5$~T~\cite{tim}) serve to
highlight the limitations of the Lifshitz-Kosevich formula at high
fields~\cite{harrison96}, where not only is there competition between two functionally
similar terms, but also the amplitudes of the quantum oscillations
become very large.

\subsection{The effect of spin-splitting} \label{g}

The energy levels of a quasiparticle in a metallic system subjected to
an applied magnetic field are defined by Landau quantisation and the
Zeeman effect, and are given by 
\beqn E=(n+\half)\frac{\hbar
eB\cos\theta}{m^*}\pm\half g^*\mu_{\rm B}B, \label{spinsplitLL} 
\eeqn
where $n$ is the Landau level index and $g^*$ is the effective
$g$-factor~\cite{review}. Increasing the $\theta$-angle reduces the separation between
Landau levels by reducing the field perpendicular to the highly
conducting planes, $B\cos\theta$. When $B\cos\theta$ is such that the
spin-up and spin-down sections of different Landau levels are degenerate
then the separation between successive energy levels is equal to
$\hbar\omega_c$. At this angle the SdH oscillations having the
fundamental frequency, $F$, will dominate, taking their maximum
amplitude. However, when $B\cos\theta$ is such that the spin-up and
spin-down sections of different Landau levels are equally spaced at
$\half\hbar\omega_c$, then the dominant oscillations will be those with
frequency 2$F$ and the amplitude of the fundamental oscillations will be
at a minimum~\cite{nicholas88}. These two situations are known as spin-maxima and spin-zeroes respectively.

It is easy to show that the conditions for spin-zeroes and spin maxima
are given by~\cite{nicholas88}  

\beqn g^*\mu_{\rm B}B=j\frac{\hbar
eB\cos\theta}{m^*}\left\{ \begin{array}{lr} j=\half, \frac{3}{2},
\frac{5}{2}... & \mbox{spin-zero~} \\ j=1, 2, 3... & \mbox{spin-max.} 
\end{array} \right. \label{szcondition} \eeqn

As has already been mentioned, this effect can be observed as a
modulation of the $\alpha$-frequency SdH oscillations when a crystal of
\cuscn~is rotated in a fixed field. \Fref{cuscnsz} shows a typical
section of such a rotation for an $h8$-sample, and several spin-zero
angles, $\theta_j$, are marked with arrows. The inset shows a plot of
$(\cos\theta_j)^{-1}$ versus $j$-index. This dataset is a summary of
a large number of spin-zero positions measured at many different values of
the azimuthal angle, at fields of 27 and 42~T, and in two different
single crystals. Using the gradient of the straight line fit shown, the
product of the effective $g$-factor and $\mu^*$, where $\mu^*=m^*/m_e$,
is found to be $g^*\mu^*_\alpha(h8)=5.22\pm0.56$.

The dataset for $d8$ is not as extensive as that for $h8$, nevertheless
a good fit is still achieved, yielding
$g^*\mu^*_\alpha(d8)=5.24\pm0.65$. The values for $h8$ and $d8$ are
identical within the errors.

\cuscnsz

The value for $g^*\mu_\alpha^*(h8)$ obtained here appears to be in good
agreement with that of Reference~\cite{wosnitza96}, obtained by fitting
three spin-zero points from de Haas-van Alphen data. In that reference
the authors assume that $g^*=2$ and that the mass obtained from the
spin-zero effect is renormalised by electron-electron interactions, but
not electron-phonon interactions, they then use the difference between
this mass and that derived from a Lifshitz-Kosevich analysis of quantum
oscillations to specify the electron-phonon coupling
constant~\cite{wosnitza96}. However, as the $g^*\mu^*$ values obtained in this manner may not be renormalised in the same way as the effective masses found from the thermodynamic variation of the quantum oscillation amplitudes or the $g$-factors obtained from electron spin resonance it is not advisable to separate $g^*$ and $\mu^*$ in this fashion.

\cuscnsplit

It is possible, using the experimentally determined value of $g^*\mu^*$,
to make a comparison of the spin and Landau level splittings. The ratio
of the splittings at $\theta=0$\deg~is given by, 
\begin{equation}
\frac{g^*\mu_{\rm B}B}{\hbar\omega_c}\equiv\frac{g^*\mu^*}{2},
\end{equation} 
using $\omega_c=eB/m^*$ and $\mu_{\rm B}=e\hbar/2m_e$.
And so for both $h8$ and $d8$ the ratio of the spin to Landau level
splitting is around 2.6, which results in the energy level diagram shown
in \Fref{cuscnsplit}. It is seen that as well as the energy splitting of
$\hbar\omega_c$ between one spin split level and its equivalent from the
next highest Landau level, there exists another splitting, of
$0.4~\hbar\omega_c$, arising from the difference between the spin-up of
one Landau level and the spin-down of the Landau level three places up.

%This splitting will become noticeable when the field and temperature are
%such that $0.4~\hbar\omega_c\geq k_{\rm B}T_{\rm D}$ (where $T_{\rm
%D}=\hbar/2\pi k_{\rm B}\tau$ is the Dingle temperature) or $k_{\rm B}T$,
%whichever is the larger, and will manifest itself in the observation of
%higher harmonics of the fundamental frequency in the Fourier transform
%spectrum of the oscillations.

\subsection{The angle-dependent magnetoresistance
oscillations} \label{amro}

\subsubsection{Boltzmann transport simulations}
\label{amrosim}

An analysis of the angular effects in \cuscn~is complicated by the
co-existence of Q1D and Q2D \fs s. The method by which the different
types of AMRO either dominate or superpose over one another is not at
all clear, depending as it does on unknowns such as the relative
effective masses and carrier densities of the quasiparticles on the Q1D
and Q2D sections. To further complicate an AMRO investigation it should
be noted that in general Lebed magic angles and Yamaji oscillations can
be analysed in very similar ways. For example, if the resistive peaks
that lie between dips caused by the Q1D Lebed magic angle effect are
accidently mistaken for Q2D Yamaji oscillations it is possible, as will
be shown later, to obtain the dimensions of a closed \fs~pocket that may
appear reasonable, but is incorrect. For this reason, when measuring
samples whose \fs~is uncharted, the Lebed magic angles and Yamaji
oscillations are best used in conjunction with other \fs~effects such as
Danner-Kang-Chaikin, or quantum oscillations, which specify exclusively the
nature of the \fs~from which they arise.

In the sample under review here, the presence of both Q1D and Q2D
sections of \fs~is not in question as it is demonstrated convincingly by
the magnetic breakdown observed in the SdH effect. However, in order to
make sense of the AMRO data measured experimentally, some method of
separating the oscillations arising from the two sections is required.
This is achieved by making detailed, semi-classical simulations of the
interplane resistivity resulting from the Q1D and Q2D \fs s. A suite of
programs were therefore developed which used Fortran (for operational
speed) to solve the equations of motion for any specified Fermi surface
and field orientation and use the results of this to find a numerical
solution to the Chambers formula (\Eref{chambers} below). This software was applied to model
Fermi surfaces and the AMRO results were seen to agree with theoretical predictions.

It is necessary here to simulate the angle-dependent effects observed in
\cuscn. To this end an equation that describes the entire \fs~of this
material throughout the first \bz~is formulated; 
\begin{eqnarray} 
E({\bf k})&=&2t_b\cos(k_xb) \nonumber \\
&& \pm\cos\left(\frac{k_xb}{2}\right)\sqrt{t_{c1}^2+t_{c2}^2
+2t_{c1}t_{c2}\cos (k_yc)} \nonumber \\
&&-2t_a\cos\left[k_za\cos(\beta-\frac{\pi}{2})
-k_ya\sin(\beta-\frac{\pi}{2})\right]\mbox{~}\label{cuscndisp} 
\end{eqnarray}
in which the conducting layers of the (ET) molecules lie in the $bc$-
(or $xy$-)plane, $\beta$ is the angle between the crystallographic $a$
and $c$ directions, and the $z$-axis lies along the interlayer
direction. The first two terms of the equation are simply the effective
dimer model that was discussed previously and which is known to
accurately describe the intralayer \fs. The last term is a tight binding
representation of the interlayer dispersion. For the purposes of the
simulations the values of $t_b$, $t_{c1}$, $t_{c2}$ and the Fermi energy
are set to be those quoted earlier, derived from a consideration of
experimental results. The transfer integral $t_a$ is set to be
0.04~meV. This is the value resulting from a preliminary
investigation of the in-plane peak effect~\cite{squit,squitnote}.

As a quasiparticle moves across the \fs~under the influence of the magnetic
field its component of velocity in a given direction will vary as
it negotiates the various \fs~contours and corrugations (its total velocity remaining perpendicular to the \fs~at all times). Considering an entire \fs~filled with orbiting quasiparticles it is not difficult to see that the combination of these varying velocity components leads to a conductivity that depends strongly on the nature of the orbits and the geometry of the \fs.

This argument is formalised in the isothermal solution to
the Boltzmann transport equation known as the Chambers formula;
\beqn \sigma_{ij}=\frac{e^2}{4\pi^3}\int d{\bf
k}^3(-\frac{df_0}{d\varepsilon})v_i({\bf
k},0)\int_{-\infty}^{0}v_j({\bf k},t)e^{t/\tau}dt,
\label{chambers} \eeqn
where $\sigma_{ij}$ is a component of the conductivity tensor,
$f_0$ is the unperturbed quasiparticle (Fermi-Dirac) distribution
function, $v_i$ and $v_j$ are velocity components and 1/$\tau$ is
the ${\bf k}$-independent scattering rate~\cite{chambers52}. This
equation represents a velocity-velocity correlation function
between the $i^{\rm th}$ component of the initial velocity,
$v_i({\bf k},0)$, integrated over all possible starting points on
the \fs, and $v_j({\bf k},t)$, the $j^{\rm th}$ component of the
velocity of a quasiparticle averaged over the duration of its
orbit. The exponential term represents the probability of a quasiparticle scattering from its trajectory so that it no longer contributes to the conductivity.

Armed with the Chambers formula and \Eref{cuscndisp}, it is now possible
to relate the way in which the program that simulates the interplane
resistivity proceeds (a discussion of the possible errors that might
creep in is left until the end): first a quasiparticle is placed at
point on the Fermi surface and its velocity components are found and
recorded by differentiating \Eref{cuscndisp}, according to ${\bf
v}=\hbar^{-1}\nabla_{\bf k}E({\bf k})$~\cite{anm}. Next, the Lorentz force
($\hbar{\rm d}{\bf k}/{\rm d}t=-e{\bf v}\times{\bf B}$~\cite{anm}) for a given
inclination of the magnetic field is allowed to act upon the
quasiparticle for a short time so that it moves to a new position on the
\fs. Here its velocity components are again recorded and the process is
repeated a large number of times so that a \fs~orbit is mapped out.

The time-integral in the Chambers formula is obtained by multiplying
each value of each velocity component by $\exp({-t/\tau})$ and adding
the like components together. The scattering time, $\tau$, is chosen to
be 3~ps in order to reflect that measured from quantum oscillations and
high frequency conductivity measurements~\cite{squit} and the whole orbit is recorded over a time
$t=8\tau$, by which point more than 99.96\% of the quasiparticles have
been scattered. The time interval, $\delta t$, between points on an
orbit was set to be quite small (never larger than $0.002t$) so that the
sum of velocity components might approximate well to the integral in the
Chambers formula. The program returns to the correct position any orbit
that has a tendency to move off the \fs~using a subroutine that exploits
the Runge-Kutta method of solving ordinary differential
equations.

The time-averaged velocity components are now multiplied by the relevant
velocity component from the start of the process, i.e.\ $t=0$, and
weighted by the density of states and the \fs~area represented by the
orbit. This routine is repeated for a large grid of starting points that
span the entire first \bz, and the results are summed. In this way the
integral over the \fs~in the Chambers formula is accomplished, and each
component of the conductivity tensor is calculated. The results are
combined to yield the interplane resistivity by inverting the conductivity
tensor; 
\begin{widetext}
\begin{equation} 
\rho_{zz}=
\frac{\sigma_{xx}\sigma_{yy}-\sigma_{xy}\sigma_{yx}}{\sigma_{xx}\sigma_{yy}\sigma_{zz}
-\sigma_{xx}\sigma_{yz}\sigma_{zy}+\sigma_{xz}\sigma_{yx}\sigma_{zy}-\sigma_{xy}\sigma_{yx}\sigma_{zz}
+\sigma_{xy}\sigma_{yz}\sigma_{xz}-\sigma_{xz}\sigma_{yy}\sigma_{zx}}.
\label{rhozz} \end{equation} \end{widetext}

This method can be used to calculate the resistivity at any value of
$\theta$, $\phi$ and $B$. The \fs~resolution chosen, i.e.\ the number of
orbits sampled, must be a compromise between the accuracy of the results
and the speed of calculation. It is found that for $\theta$-angles away
from 90\deg~a grid of 100$\times$100 starting points is sufficient to
successfully simulate the resistivity. However, close to 90\deg~the
orbits are rapidly changing with $\theta$, and the interplane
resistivity is dominated by a few small, closely-spaced orbits. In this
case it is necessary to greatly increase the \fs~resolution, which in
turn greatly lengthens the duration of the simulation.

\Q1dsim

In performing these simulations the interest lies in their ability to
reproduce the Lebed magic angle effect and the Yamaji oscillations, as
it is these phenomena that need to be distinguished from one another.
Less important are the Danner-Kang-Chaikin oscillations and the third angular
effect. Although they too are reproduced by the Chambers formula, an
analysis of these effects does not yield a great deal of useful
information. That said, the in-plane peak effect, which is intimately related to
both the Danner-Kang-Chaikin and third angular oscillations, is of great
interest, but will be dealt with in a different manner in~\Sref{squit}.

%A selection of results using the simulation method described above
%have already been seen as figures illustrating the various
%magnetoresistance effects in \Cref{mr}.

Simulations of the angle-dependence of the interplane resistivity at
42~T and several values of the azimuthal angle, $\phi$, for the Q1D
sections of \fs~are shown in \Fref{q1dsim}. At $\phi=\pm90$\deg, which
corresponds to the magnetic field lying parallel to the Q1D sheets, the
Lebed magic angle effect can be clearly seen as dips in the
magnetoresistance. As the $\phi$-angle is changed the frequency of the
dips also changes. At low $\phi$-angles the amplitude of the dips
drops, and at $\phi=0$\deg~they are no longer observed. The
Danner-Chaikin oscillations are seen as smaller features near
$\theta=90$\deg~at low azimuthal angles.

\q1dsimres

The validity of these simulations can be checked by calculating the
frequency in $(\tan \theta)^{-1}$, $1/\chi$, of the Lebed magic angle dips for each value of the azimuthal angle. In this way the $\phi$-dependence of $\chi$ can be
fitted to the equation
\begin{equation}
\chi(\phi)=\frac{\chi_0}{\cos(\phi-\phi_0)}, \label{lebedphidep}
\end{equation}
where $\phi_0$ corresponds to the magnetic field lying parallel to
the Q1D sheets~\cite{caulfield95}. The results of such an analysis are shown
in \Fref{q1dsimres}. From the fit the value of $\chi_0$, which is equal
to the $c$ lattice parameter divided by the interlayer distance,
$d_\perp$, is found to be $0.8658\pm0.0005$. This can be compared with
the value of $c/d_\perp=0.861\pm0.001$ obtained from X-ray scattering
measurements~\cite{urayama88}. 

The inset to \Fref{q1dsimres} shows the results of analysing the Lebed magic angles if they were mistakenly taken for Yamaji oscillations. The dotted line is a fit to the equation
\beqn k_\parallel^{\rm max}
(\phi)=[k_a^2\cos^2(\phi-\xi)+k_b^2\sin^2(\phi-\xi)]^{1/2},
\label{kpara} \eeqn
which is valid for a \fs~with an elliptical cross-section and where $k_\parallel^{\rm max}$ is the maximum in-plane Fermi wavevector projected on the plane of rotation of the field and is found from the frequency of the Yamaji oscillations. $k_a$ and $k_b$ are the major and minor semi-axes of the Q2D \fs~pocket respectively~\cite{house96}. This fit suggests the existence of an elliptical Q2D pocket with major and minor axes of 2.37~nm$^{-1}$ and 0.29~nm$^{-1}$ respectively. If these were experimental results then it is easy to see that in the absence of any other evidence such a \fs~pocket might seem quite reasonable. However, the mistake becomes apparent when the fundamental frequency of the quantum oscillations that would be expected from a closed pocket of this size, 227~T, is compared with the experimentally determined value
of 599~T.

\aQ2dsim

\Fref{q2dsim} shows the simulated interplane magnetoresistance that
arises from the Q2D closed \fs~pockets at 42~T and various $\phi$-angles. It is
seen that the traces are dominated by the peaks of the Yamaji
oscillations. $k_\parallel^{\rm max}$ can be extracted from the
frequency of the oscillations at each $\phi$-angle, and the result of
fitting this to \Eref{kpara} is shown in \Fref{aq2dsimres}. The
resulting Q2D pocket has a major axis, $k_b$, of
$2.476\pm0.001$~nm$^{-1}$, and a minor axis, $k_c$, of
$0.733\pm0.002$~nm$^{-1}$. This would give rise to quantum oscillations
with a fundamental frequency of 598~T, which is in agreement with the
value measured from the SdH effect.

\aq2dsimres

It is now possible to mark the differences expected between the shape of
the Q2D pocket that results from the correct analysis of the Yamaji
angles, and that from the mistaken identification of the Lebed magic
angles. The obvious difference is that the pockets are perpendicular to
each other, with the long axis of the true Q2D pocket lying along the
$\phi=0$\deg~direction. If the samples used in the experiments had been
oriented by optical measurements then this would be sufficient to
distinguish the AMROs. However, this is not the case. The major axes of
the two alleged pockets are similar to each other, and an experimental
error is likely to encompass them both. Thus it is to the minor axis
that one must look to separate the two AMRO effects.

All in all the simulations agree very well with the experimental results
of both X-ray scattering and the SdH effect. Nevertheless, it is
worthwhile to look more closely at the various errors that might be
introduced into the simulation process along the way. The first, most
general problem to be addressed is that the simulations are
semi-classical, and take no account of the quantum oscillations and,
more importantly, the magnetic breakdown. As the perpendicular field is
increased to high magnitudes, the experimentally measured AMRO will become affected by magnetic breakdown, as more and more Q1D
carriers tunnel through the energy gap and become Q2D carriers.
Eventually the system will resemble one large Q2D \fs~pocket whose
cross-section in the highly conducting planes is the $\beta$-orbit.
However, AMRO tend to be most concentrated near to $\theta=90$\deg. In
fact, the actual experiments were performed at 42~T, and it will be seen
that almost all the important AMRO features occur at $\theta$-angles of around
70\deg~or higher. The perpendicular magnetic field at $B=42$~T,
$\theta=70$\deg~is such that for the $d8$-sample less than one
quasiparticle in fourteen has sufficient energy to bridge the gap
between the \fs s. For $h8$ this value is less than one in fifty, and
the probability of breakdown for both types of sample decreases towards zero as $\theta$ approaches 90\deg (whether or not the probability actually reaches zero at $\theta$=90\deg~depends on the relative sizes of $t_a$ and $E_g$). Thus for the current situation the
magnetic breakdown is only a minor consideration.

The most likely entry point for errors to make their way into the
calculations is via the values chosen to represent the various physical
parameters. It has already been mentioned that the values chosen for
$t_b$, $t_c$ and $E_{\rm F}$ used in conjunction with the effective
dimer model reproduce the measured \fs~very well, so attention is turned
to the other parameters, namely $E_g$ and $t_a$. The value of 7.8~meV
chosen for $E_g$ is derived from a measured value of the magnetic
breakdown, which has a large error associated with
it~\cite{harrison96mb}. However, a quick glance at how such an error
propagates reveals it to be relatively unimportant: $E_g$ represents the
gap in $k$-space between the Q1D and Q2D sections of \fs, and an order of
magnitude estimate of this gap in terms of wavevector, $\Delta k$, is
given by $\Delta k/k_{\rm F}\sim E_g/E_{\rm F}$~\cite{shoenberg}. Using
estimates for $k_{\rm F}$ and \Ef~\cite{sasaki91}, it is found that
$\Delta k\sim 0.4$~nm$^{-1}$. As the area of the Q2D pocket is well
defined, any error on the size of the gap would lead to errors in $k_b$
and $k_c$, the axes that define the pocket. A generous error on $k_b$ is
$\pm0.2$~nm$^{-1}$, or half of $\Delta k$, which represents what would
happen if the energy gap were allowed to be zero. Fixing the area, this
leads to a error of around $\pm0.1$~nm$^{-1}$ on $k_c$. Even with such
an uncertainty on the magnitude of the minor axis, it would still be
possible to distinguish between the results arising from the Yamaji
oscillations and those from the Lebed magic angles.

The value used to represent $t_a$, the transfer integral along the
crystallographic $a$ direction, is based on a preliminary analysis of
the in-plane peak effect~\cite{squit}. As it is quite small, 0.04~meV, it is
likely to have associated with it a significant relative error. However,
the positions of the AMRO features arising from the Yamaji and Lebed
effects are unaffected by the magnitude of the $t_a$ parameter. In the
current situation the amplitudes of the oscillations are of little
concern, thus, for the moment, neither is the precise value $t_a$.

\sasamro

\subsubsection{Experiments} \label{expamro}

\ah8results

\Fref{sasamro} shows a selection of the measured angle-dependences of
$h8$ \cuscn, at various values of the azimuthal angle, in a field of
42~T and at temperatures around 500~mK. In order to analyse such
angle-dependences, the position of each AMRO peak and dip is recorded.
The frequency in $(\tan\theta)^{-1}$ of the peaks and dips at each
$\phi$-angle is then found for each sample, and the results are compared
to those obtained from the simulated resistance. The AMRO arising from
the Q1D and Q2D \fs s are thus identified and the measured $\phi$-angle
can be calibrated so that $\phi=0$\deg~is perpendicular to Q1D sheets.
The frequencies of the dips arising from the Q1D Lebed magic angle
effect for each sample are combined and fitted to \Eref{lebedphidep}.
The result is shown in the left hand side of \Fref{ah8res}. The fit is good and it is found that $\chi_0(h8)=c/d_\perp=0.89\pm0.10$, which is in reasonable
agreement with value of $0.861\pm0.001$ found from X-ray
scattering~\cite{urayama88}.

The results of calculating $k_\parallel^{\rm max}(\phi)$ for the
resistance peaks arising from the Q2D Yamaji oscillations for each
sample are also combined, and these data, together with the curve
obtained by fitting to \Eref{kpara}, are shown in the right hand side of \Fref{ah8res}. 

It is seen that almost all the $k_\parallel^{\rm max}(\phi)$ data are
concentrated around the region where the magnetic field is roughly
perpendicular to the flattish portion of the Q2D pocket --- at the other
$\phi$-angles the resistance is dominated by the Lebed effect. This
means that the major axis, $k_b$, is ill-defined and it is necessary
when performing the fit to fix the area of the pocket so that it
reproduces the measured fundamental frequency of the SdH oscillations.
It can be seen from the figure that this fit is reasonable; and the
results obtained are $k_c(h8)=0.80\pm0.05$~nm$^{-1}$ and
$k_b(h8)=2.28\pm0.15$~nm$^{-1}$.

\ch8resall
\3rdang
\bd8results

It is illustrative to calculate $k_\parallel^{\rm max}(\phi)$ for all
the $h8$ data (Lebed angles and Yamaji oscillations) and display the
results on the same polar plot. In this way it is easy to see which
\fs~section dominates the resistance at a given azimuthal angle. This
plot is shown in \Fref{h8resall}. The hollow symbols are the Lebed magic
angle dips, and the solid symbols are the Yamaji peaks. The dotted lines
are the fitted curves from the previous two figures. It is clear that
the Q1D \fs~sheets dominate the angle-dependent magnetoresistance when
the field is roughly perpendicular to the sheets, and the Q2D pockets
dominate when the field is roughly perpendicular to their flatter edges.

In order to shed light on this behaviour, \Fref{3rdang} shows the result
of simulating the interlayer resistance at fixed field of 42~T and a  $\theta$-angle of
90\deg~over the whole range of azimuthal angle, for both the Q1D and Q2D
\fs~sections. This is just the simulated third angle effect
plotted in polar coordinates, and is chosen to be representative of the
magnitude of the resistance at 42~T in the angular region (70\deg~--90\deg) over
which AMRO are observed. By qualitatively comparing
Figures~{\ref{h8resall}} and~{\ref{3rdang}}, it is seen that the range
of azimuthal angles over which the Q1D \fs~dominates the measured
magnetoresistance is similar to that over which the simulated
$R_{zz}(\theta=90^\circ)$ due to the Q1D sheets is lower than that due
to the Q2D pockets. The inverse is true for the range of angles where
the measured magnetoresistance is dominated by Q2D AMRO effects. Given
that the simulations have already been shown to be trustworthy by
reliably reproducing experimental data, it appears that the resistances
from the two sections of \fs~combine in similar way to resistors in
parallel, i.e.\ the overall resistance of the system at a given
$\phi$-angle is dominated by the section of \fs~that takes the lowest
resistance at that angle. However, the situation away from $\theta=90^\circ$ is not quite as simple
as the parallel resistors scenario, as it is found that adding the
simulated angle-dependent resistances for each \fs~section using $R_{\rm
total}^{-1}=R_{\rm Q1D}^{-1}+R_{\rm Q2D}^{-1}$ does not successfully
reproduce the experimental results.

Similar AMRO effects were measured in the $d8$ sample. The left hand side of \Fref{bd8res}
shows the azimuthal angle-dependence of the frequency of the Lebed magic
angle dips. The fit
to \Eref{lebedphidep} is reasonable but not nearly as good as that for the $h8$ sample. The value of
$\chi_0(d8)=c/d_\perp$ is found to be $0.70\pm0.15$. This can be
compared to the value of $0.862\pm0.001$ found from X-ray scattering
measurements of the deuterated salt~\cite{saito91}. It is not entirely clear why the results are not as successful for the $d8$ sample as for the $h8$, however 
it is seen that there are
nearly half the number of data points in the $d8$ fit than the $h8$, and
further, the density of points in the regions around
$\phi=90^\circ$, where the $\chi_0$ parameter is best defined, is
much lower in the case of the $d8$ fit.

The right hand side of \Fref{bd8res} shows the azimuthal angle-dependence of
$k_\parallel^{\rm max}$ as calculated from the frequency of the Yamaji
oscillations. In order to fit the data to \Eref{kpara}, the area is
again constrained to produce quantum oscillations of the correct
frequency. The axes of the pocket are thus found to be;
$k_c(d8)=0.83\pm0.13$~nm$^{-1}$ and $k_b(d8)=2.19\pm0.35$~nm$^{-1}$.

\begin{table} \caption{Comparison of results derived from the AMRO of
$h8$ and $d8$ \cuscn, and from the semi-classical transport
simulations.} \centering \begin{tabular}{lccc} \hline & $\chi_0$ & $k_b$
(nm$^{-1}$) & $k_c$ (nm$^{-1}$)\\ 
\hline 
h8 & $0.89\pm0.10$ & $2.28\pm0.15$ & $0.80\pm0.05$\\ 
d8 & $0.70\pm0.15$ & $2.19\pm0.35$ & $0.83\pm0.13$\\ 
simulations & $0.8658\pm0.0005$ & $2.476\pm0.001$ & $0.733\pm0.002$\\ 
\hline 
\label{amrodat} 
\end{tabular} 
\end{table}

A comparison of the results of analysing the AMRO measured in the $h8$
and $d8$ samples, and those simulated using the Chambers Formula, is
shown in \Tref{amrodat}. It is seen that the results for $h8$ and $d8$
agree with each other to within the error ranges. It is also seen that there is reasonable correlation
between the experimentally determined values and those from the
simulations. This implies that the parameters used in the simulation
program are good approximations to the real values, and that it is
possible to explain the AMRO in terms of purely semi-classical effects.
\squitorbs
\squitwidth
\subsection{Characterising the interlayer transport}
\label{squit}

For a system with a three-dimensional \fs, a series of quasiparticle orbits are possible in the presence of an exactly in-plane magnetic field, many of which are very good at averaging the interlayer velocity towards zero (see \Fref{squitorbs}). This is the origin of the in-plane peak effect in \rhozz, and suggests a coherent nature to the interlayer transport~\cite{mckenzie98,hanasaki98}.

For highly anisotropic materials, the angular width, $2\Delta$, of the in-plane peak, when measured
in radians, can be approximated by $2v_\perp^{\rm max}/v_\parallel$, where $v_\perp^{\rm max}$ is the maximum of the out-of-plane component of the quasiparticle velocity and $v_\parallel$ is the in-plane component parallel to the plane of rotation of the magnetic field~\cite{squit}. The $\phi$-dependence of $v_\parallel$ can be calculated for \cuscn~using the dispersion relation in \Eref{cuscndisp} and the
in-plane \fs~parameters discussed in \Sref{fs}. $v_\perp$ is given
by $\hbar^{-1}\partial E/\partial k_z$, and so $v_\perp^{\rm max}$ is a
constant equal to $2t_aa\cos(\beta-\pi/2)/\hbar$. In this way 2$\Delta$
is calculated, with the value of $t_a$ left as the only adjustable
parameter. This value is determined by comparing the results of the
calculation with the experimentally derived values for the width of the in-plane peak. This is the same method used in Reference~\cite{squit}, however in the present case the monoclinic structure of \cuscn~is taken into account, and hence slightly different results are achieved. 

The result for the $h8$ sample is shown in the left hand side of \Fref{squitwidth}. Here the points are the experimental data for two
$h8$ samples, and the solid lines are the results of the calculations
obtained by setting $t_a(h8)=0.065\pm0.007$~meV. The continuous curve arises
from closed orbits on the Q2D FS pocket, which are possible when the
magnetic field is directed along any $\phi$-angle. The closed loops
correspond to orbits about the Q1D FS sheets, which are only possible
when the field is directed along a limited range of $\phi$. Away from
this $\phi$-range the data follows the continuous curve fairly well and
agrees with the $\phi$-calibration found in the previous section. Around
$\phi=0$\deg, 180\deg, and 360\deg~the width of the peak can be governed
by any of the three sets of closed orbits possible; those on the broadly
curved, convex region of Q1D sheets; those on the pointed, concave
region located at the \bz~boundary; or those on the Q2D pocket.
The way in which these orbits will combine to produce the in-plane peak is not entirely clear, but it might be expected that the conductivities of each orbit sum to produce the total conductivity, as is the case in the Chambers formula. 
To the first approximation the interlayer resistivity is found by
inverting the interlayer conductivity and so,
\begin{equation}
\frac{1}{\rho_{zz}}\approx \sigma_{zz}=
\sigma_{zz_1}+\sigma_{zz_2}+\sigma_{zz_3}+~...\approx
\frac{1}{\rho_{zz_1}}+\frac{1}{\rho_{zz_2}}+\frac{1}{\rho_{zz_3}}+~...
\label{parallelres}
\end{equation}
i.e.\ the resistivity contributions from each orbit combine like
resistors in parallel, and it is the path with the smallest
resistivity that dominates the total resistance. Thus, in terms of
the in-plane peak effect it is likely that the orbits that are the least efficient at averaging the interlayer velocity towards zero will dominate the resistance.

The results for the $d8$ sample are shown in the right hand side of \Fref{squitwidth}. The
best agreement between calculations and experiment was found by setting
$t_a(d8)=0.045\pm0.005$~meV.

\section{Conclusions} \label{conclusions}

In summary, several physical properties have been measured for both 
hydrogenated and deuterated \cuscn. No disparity has been
found in the size and shape of the \fs s of the two isotopes, their
effective masses, their scattering rates or their energy level
structures. The only discernable difference found was in the interlayer transfer integral, which appeared lower for the deuterated salt.

The size of the interlayer warping is determined by the transfer
integral in this direction. An increased warping means that the \fs~will
be less able to nest. Thus, if indeed the superconductivity in this material is aided by nestability~\cite{schmalian98,kondo98,kuroki99} then the higher transfer integral in the
$h8$-salt would help to explain its lower superconducting transition
temperature.

Further, it has been shown that the measured angle-dependent
magnetoresistance oscillations can be reproduced via purely
semi-classical, Boltzmann transport considerations. The observed peak in
the resistance in the presence of a nearly in-plane magnetic field
suggests that at low temperatures and ambient pressure the \fs~of \cuscn~is a three-dimensional object that extends throughout reciprocal space.
Thus it is not necessary to invoke non-Fermi liquid effects in order to
describe the angle-dependent interlayer transport in this material.

This work is supported by EPSRC (UK).
NHMFL is supported by the
US Department of Energy (DoE), the National
Science Foundation and the State of Florida.
Work at Argonne is sponsored
by the DoE, Office of Basic Energy Sciences,
Division of Materials Science under contract number
W-31-109-ENG-38.

\end{sloppypar}
%\bibliography{ABC}

\begin{thebibliography}{10}
\bibitem{squitnote}
\Eref{cuscndisp} differs from the dispersion relation quoted in~\cite{squit}
  because in the present paper the monoclinic structure of
  \cuscn~(\mbox{$\beta=110.30 \pm 0.03$\deg}~\cite{urayama88}) is taken into
  account.
\end{thebibliography}

\end{document}